\documentclass[journal,twocolumn,twoside]{IEEEtran}
\usepackage{array}
\usepackage{color}
\makeatletter
\def\blfootnote{\gdef\@thefnmark{}\@footnotetext}
\makeatother
\RequirePackage{filecontents}
\usepackage{cite}
\usepackage{graphicx}
\usepackage{psfrag}
\usepackage{paralist}
 \usepackage{multirow}
 \usepackage[table,xcdraw]{xcolor}
\usepackage[acronym]{glossaries}
\usepackage{amsmath,stackrel}
\usepackage{amsfonts}
\usepackage{mathalfa}
\usepackage{amssymb,mathtools}
\usepackage{amscd}
\usepackage{tikz}
\usepackage{caption}
\usepackage{subcaption}
\usepackage{verbatim}
\usepackage{enumerate}                     
\usepackage{amsthm}
\usepackage{eucal}
\usepackage{algorithmic}
\usepackage{algorithm}
\usepackage[nolist]{acronym}
\usepackage[normalem]{ulem}
\usepackage{algorithm,algorithmic} 
\usepackage[capitalise]{cleveref}
\Crefname{equation}{Eq.\!}{Eqs.\!}
\Crefname{figure}{Fig.\!}{Figs.\!}
\Crefname{tabular}{Tab.\!}{Tabs.\!}
\Crefname{section}{Section\!}{Sections.\!}

\begin{document}
\begin{acronym}
\acro{lora}[LoRa]{low power long range}
\acro{lorawan}[LoRaWAN]{Low power, Long Range Wide Area Network}
\acro{lpwan}[LPWAN]{Low Power Wide Area Network}
\acro{LDPC}{low-density parity-check}
\acro{ed}[ED]{end device}
\acro{ns}[NS]{network server}
\acro{gw}[GW]{gateway}
\acro{fsk}[FSK]{frequency shift key}
\acro{sf}[SF]{spreading factor}
\acro{ISM}{Industrial, Scientific and Medical}
\acro{ETSI}{European telecommunications standards institute}
\acro{CPS}{cyber-physical systems}
\acro{CPScapital}[CPS]{Cyber-physical systems}
\acro{AC}{address coding}
\acro{ACF}{autocorrelation function}
\acro{ACR}{autocorrelation receiver}
\acro{ADC}{analog-to-digital converter}
\acrodef{aic}[AIC]{Analog-to-Information Converter}     
\acro{AIC}[AIC]{Akaike information criterion}
\acro{aric}[ARIC]{asymmetric restricted isometry constant}
\acro{arip}[ARIP]{asymmetric restricted isometry property}

\acro{ARQ}{automatic repeat request}
\acro{AUB}{asymptotic union bound}
\acrodef{awgn}[AWGN]{Additive White Gaussian Noise}     
\acro{AWGN}{additive white Gaussian noise}
\acro{waric}[AWRICs]{asymmetric weak restricted isometry constants}
\acro{warip}[AWRIP]{asymmetric weak restricted isometry property}
\acro{BCH}{Bose, Chaudhuri, and Hocquenghem}        
\acro{BCHC}[BCHSC]{BCH based source coding}
\acro{BEP}{bit error probability}
\acro{BER}{bit error rate}
\acro{BFC}{block fading channel}
\acro{BG}[BG]{Bernoulli-Gaussian}
\acro{BGG}{Bernoulli-Generalized Gaussian}
\acro{BPAM}{binary pulse amplitude modulation}
\acro{BPDN}{Basis Pursuit Denoising}
\acro{BPPM}{binary pulse position modulation}
\acro{BPSK}{binary phase shift keying}
\acro{BPZF}{bandpass zonal filter}
\acro{BSC}{binary symmetric channels}              
\acro{BU}[BU]{Bernoulli-uniform}
\acro{bpcu}{bit per channel use}
\acro{BICM}{bit-interleaved coded modulation}

\acro{CM}{coded modulation}
\acro{CDF}{cumulative distribution function}
\acro{CCDF}{complementary cumulative distribution function}
\acro{CD}{cooperative diversity}

\acro{CDMA}{code division multiple access}
\acro{ch.f.}{characteristic function}
\acro{CIR}{channel impulse response}
\acro{cosamp}[CoSaMP]{compressive sampling matching pursuit}
\acro{CR}{cognitive radio}
\acrodef{cscapital}[CS]{Compressed sensing}
\acro{CS}[CS]{compressed sensing}
\acro{CSI}{channel state information}
\acro{CCDM}{constant composition distribution matcher}
\acro{coms}[SpaDCoM]{sparse-dense with constant composition based coded modulation for \ac{FSO} communication}

\acro{DAA}{detect and avoid}
\acro{DAB}{digital audio broadcasting}
\acro{DCT}{discrete cosine transform}
\acro{dft}[DFT]{discrete Fourier transform}
\acro{DR}{distortion-rate}
\acro{DS}{direct sequence}
\acro{DS-SS}{direct-sequence spread-spectrum}
\acro{DTR}{differential transmitted-reference}
\acro{DVB-H}{digital video broadcasting\,--\,handheld}
\acro{DVB-T}{digital video broadcasting\,--\,terrestrial}
\acro{DVB-S2}{second generation digital video broadcasting over satellite}
\acro{DC}{difference of convex}
\acro{DMC}{discrete memoryless channel}
\acro{DM}{distribution matching}
\acro{PBDM}{partition-based distribution matching}

\acro{ECC}{European Community Commission}
\acro{EED}[EED]{exact eigenvalues distribution}
\acro{ELP}{equivalent low-pass}

\acro{FSO}{free-space optical}
\acro{FC}[FC]{fusion center}
\acro{FCC}{Federal Communications Commission}
\acro{FEC}{forward error correction}
\acro{FFT}{fast Fourier transform}
\acro{FH}{frequency-hopping}
\acro{FH-SS}{frequency-hopping spread-spectrum}
\acrodef{FS}{Frame synchronization}
\acro{FSK}{Frequency Shift Key}
\acro{FSsmall}[FS]{frame synchronization}

\acro{GA}{Gaussian approximation}
\acro{GF}{Galois field }
\acro{GG}{Generalized-Gaussian}
\acro{GIC}[GIC]{generalized information criterion}
\acro{GLRT}{generalized likelihood ratio test}
\acro{GPS}{Global Positioning System}
\acro{GMI}{generalized mutual information}

\acro{HAP}{high altitude platform}

\acro{IDR}{information distortion-rate}
\acro{IFFT}{inverse fast Fourier transform}
\acro{iht}[IHT]{iterative hard thresholding}
\acro{i.i.d.}{independent, identically distributed}
\acro{IoT}{Internet of Things}                      
\acro{IR}{impulse radio}
\acro{lric}[LRIC]{lower restricted isometry constant}
\acro{lrict}[LRICt]{lower restricted isometry constant threshold}
\acro{ISI}{intersymbol interference}
\acro{IM/DD}{intensity modulation/direct detection}

\acro{LEO}{low earth orbit}
\acro{LF}{likelihood function}
\acro{LLF}{log-likelihood function}
\acro{LLR}{log-likelihood ratio}
\acro{LLRT}{log-likelihood ratio test}
\acro{LOS}{line-of-sight}
\acro{LRT}{likelihood ratio test}
\acro{wlric}[LWRIC]{lower weak restricted isometry constant}
\acro{wlrict}[LWRICt]{LWRIC threshold}
\acro{LUT}{lookup table}

\acro{MB}{multiband}
\acro{MC}{multicarrier}
\acro{MDS}{mixed distributed source}
\acro{MF}{matched filter}
\acro{m.g.f.}{moment generating function}
\acro{MI}{mutual information}
\acro{MIMO}{multiple-input multiple-output}
\acro{MISO}{multiple-input single-output}
\acrodef{maxs}[MJSO]{maximum joint support cardinality}                                           
\acro{ML}[ML]{maximum likelihood}
\acro{MMSE}{minimum mean-square error}
\acro{MMV}{multiple measurement vectors}
\acrodef{MOS}{model order selection}
\acro{M-PSK}[${M}$-PSK]{$M$-ary phase shift keying}                        
\acro{M-QAM}[$M$-QAM]{$M$-ary quadrature amplitude modulation}             
\acro{MRC}{maximal ratio combiner}                  
\acro{maxs}[MSO]{maximum sparsity order}                                    
\acro{M2M}{machine to machine}                                                
\acro{MUI}{multi-user interference}
\acro{MI}{mutual information}
\acro{MPDM}{Multiset-Partition Distribution Matching}      
\acro{MLC}{multilevel coding}
\acro{MSD}{multistage decoding}

\acro{NB}{narrowband}
\acro{NBI}{narrowband interference}
\acro{NLA}{nonlinear sparse approximation}
\acro{NLOS}{non-line-of-sight}
\acro{NTIA}{National Telecommunications and Information Administration}
\acro{NGMI}{normalized generalized mutual information}

\acro{OC}{optimum combining}
\acro{ODE}{operational distortion-energy}
\acro{ODR}{operational distortion-rate}
\acro{OFDM}{orthogonal frequency-division multiplexing}
\acro{omp}[OMP]{orthogonal matching pursuit}
\acro{OSMP}[OSMP]{orthogonal subspace matching pursuit}
\acro{OOK}{on-off keying}

\acro{PAM}{pulse amplitude modulation}
\acro{PAR}{peak-to-average ratio}
\acrodef{pdf}[PDF]{probability density function}                      
\acro{PDF}{probability density function}
\acrodef{p.d.f.}[PDF]{probability distribution function}
\acro{PDP}{power dispersion profile}
\acro{PMF}{probability mass function}                              
\acrodef{p.m.f.}[PMF]{probability mass function}
\acro{PN}{pseudo-noise}
\acro{PPM}{pulse position modulation}
\acro{PRake}{Partial Rake}
\acro{PSD}{power spectral density}
\acro{PSEP}{pairwise synchronization error probability}
\acro{PSK}{phase shift keying}
\acro{8-PSK}[$8$-PSK]{$8$-phase shift keying}
\acro{PS}{probabilistic shaping}
\acro{PAS}{probabilistic amplitude shaping}
\acro{PEP}{pairwise error probability}
\acro{MPAM}[$M$-PAM]{$M$-ary pulse amplitude modulation}

\acro{QAM}{quadrature amplitude modulation}
\acro{QPSK}{quadrature phase shift keying}

\acro{RD}[RD]{raw data}
\acro{RDL}{"random data limit"}
\acro{ric}[RIC]{restricted isometry constant}
\acro{rict}[RICt]{restricted isometry constant threshold}
\acro{rip}[RIP]{restricted isometry property}
\acro{ROC}{receiver operating characteristic}
\acro{rq}[RQ]{Raleigh quotient}
\acro{RS}[RS]{Reed-Solomon}
\acro{RSC}[RSSC]{RS based source coding}
\acro{r.v.}{random variable}                                
\acro{R.V.}{random vector}
\acro{RF}{radio frequency}

\acro{SA}[SA-Music]{subspace-augmented MUSIC with OSMP}
\acro{SCBSES}[SCBSES]{Source Compression Based Syndrome Encoding Scheme}
\acro{SCM}{sample covariance matrix}
\acro{SEP}{symbol error probability}
\acro{SG}[SG]{sparse-land Gaussian model}
\acro{SIMO}{single-input multiple-output}
\acro{SINR}{signal-to-interference plus noise ratio}
\acro{SIR}{signal-to-interference ratio}
\acro{SISO}{single-input single-output}
\acro{SMV}{single measurement vector}
\acro{SNR}[\textrm{SNR}]{signal-to-noise ratio} 
\acro{sp}[SP]{subspace pursuit}
\acro{SS}{spread spectrum}
\acro{SW}{sync word}
\acro{SER}{symbol error rate}
\acro{SMD}{symbol-metric decoder}
\acro{BMD}{bit-metric decoder}
\acro{SDT}{sparse-dense transmission}

\acro{TH}{time-hopping}
\acro{ToA}{time-of-arrival}
\acro{TR}{transmitted-reference}
\acro{TW}{Tracy-Widom}
\acro{TWDT}{TW Distribution Tail}

\acro{UAV}{unmanned aerial vehicle}
\acro{uric}[URIC]{upper restricted isometry constant}
\acro{urict}[URICt]{upper restricted isometry constant threshold}
\acro{UWB}{ultrawide band}
\acro{UWBcap}[UWB]{Ultrawide band}            
\acro{wuric}[UWRIC]{upper weak restricted isometry constant}
\acro{wurict}[UWRICt]{UWRIC threshold}

\acro{WiM}[WiM]{weigh-in-motion}
\acro{WLAN}{wireless local area network}
\acro{wm}[WM]{Wishart matrix}                               
\acroplural{wm}[WM]{Wishart matrices}
\acro{WMAN}{wireless metropolitan area network}
\acro{WPAN}{wireless personal area network}
\acro{wric}[WRIC]{weak restricted isometry constant}
\acro{wrict}[WRICt]{weak restricted isometry constant thresholds}
\acro{wrip}[WRIP]{weak restricted isometry property}
\acro{WSN}{wireless sensor network}                        
\acro{WSS}{wide-sense stationary}

\acro{FER}{frame error rate}

\end{acronym}

\def\I{\mathbb{I}}	
\def\p{{\mathbf p}}
\def\pj{p_{j}}
\def\po{\dot{\mathbf{p}}}
\def\poo{\ddot{\mathbf{p}}}
\def\poj{\dot{{p}}_{j}}
\def\SNR{\mathrm{SNR}}
\def\J{\mathbf{J}}
\def\u{{\mathbf u}}
\def\Ip{{\mathbb{I}}_{\Delta}(\p|g)}
\def\Iu{{\mathbb{I}}_{\Delta}(\u|g)}
\def\Hp{\mathbb{H}(X_{\text p})}
\def\rc{c}
\def\d{\mathrm{d}}
\def\rb{R_{\text{backoff}}}
\def\X{X}
\def\Y{Y}
\def\x{x}
\def\y{y}
\def\W{W}
\def\w{w}
\def\H{G}
\def\h{h}
\def\n{n} 
\def\fx{f_{\X}}

\def\Xi{X_i}
\def\Xu{X_u}
\def\Xs{X_s}
\def\fxi{f_{\Xi}}
\def\fxs{f_{\Xs}}
\def\fxu{f_{\Xu}}
\def\xs{x_{s}}
\def\m{j}
\def\pm{p_{\m}}
\def\aj{a_{\m}}
\def\a{{\mathbf a}}
\def\pw{P}
\def\rpam{C}
\def\rts{{R}_{\text{SDT}}}
\def\cts{{C}_{\text{SDT}}}
\def\ror{R_\text{SDOR}}

\title{Adaptive Coded Modulation for IM/DD Free-Space Optical Backhauling: A Probabilistic Shaping Approach}
\author{Ahmed~Elzanaty,~\IEEEmembership{Member,~IEEE}, and Mohamed-Slim Alouini,~\IEEEmembership{Fellow,~IEEE} 
\thanks{The authors are with the Computer, Electrical, and Mathematical Science and Engineering (CEMSE) Division, King Abdullah University of Science and Technology (KAUST), Thuwal, Makkah Province, Saudi Arabia. (e-mail: \{ahmed.elzanaty,slim.alouini\}@kaust.edu.sa).}}
\markboth{Accepted For Publication in IEEE Transactions on Communications}{Elzanaty {\MakeLowercase{\textit{et al.}}}: Adaptive Coded Modulation for IM/DD Free-Space Optical Backhauling: A Probabilistic Shaping Approach}
\maketitle
\begin{abstract}
	In this paper, we propose a practical adaptive coding modulation scheme to approach the capacity of \ac{FSO} channels with \acl{IM/DD} based on probabilistic shaping. The encoder efficiently adapts the transmission rate to the \acl{SNR}, accounting for the fading induced by the atmospheric turbulence. The transponder can support an arbitrarily large number of transmission modes using a low complexity channel encoder with a small set of supported rates. Hence, it can provide a solution for \ac{FSO} backhauling in terrestrial and satellite communication systems to achieve higher spectral efficiency. We propose two algorithms to determine the capacity and capacity-achieving distribution of the scheme with unipolar \ac{MPAM} signaling. Then, the signal constellation is probabilistically shaped according to the optimal distribution, and the shaped signal is channel encoded by an efficient binary \acl{FEC} scheme.  Extensive numerical results and simulations are provided to evaluate the performance.  The proposed scheme yields a rate close to the tightest lower bound on the capacity of \ac{FSO} channels. For instance, the coded modulator operates within $0.2$~dB from the \ac{MPAM} capacity, and it outperforms uniform signaling with more than $1.7$~dB, at a transmission rate of $3$~bits per channel use. 
	\acresetall
\end{abstract}
%
%
\section{Introduction}
The availability of efficient backhauling for terrestrial and satellite communication systems facilitates the design and widespread of these networks. Dense placement of small cells is required to permit high rates in cellular networks, leading to high-cost backhauling for the increased number of cells. Also, to expedite ubiquitous wireless coverage, efficient backhauling between satellites or unmanned flying platforms and core networks are essential. The main possibilities for backhauling are wired links, e.g., fiber optics, and wireless links, e.g., microwave and free-space optics\cite{AlzShakYanAlouini:18}. Optical fiber links are considered as a suitable solution; however, the high-cost involved in deploying cables restricts their usage in remote areas. Microwave links are another alternative solution; nevertheless, the spectrum is congested, and the available bandwidth does not support the required high rate \cite{KhaUysal:14}. On the contrary, \ac{FSO}  communication techniques can provide long-distance high-speed links, at lower cost\cite{KhaUysal:14}. Therefore, \ac{FSO}  backhauling is considered as a promising candidate for beyond-5G terrestrial and next-generation satellite communication systems\cite{AlzShakYanAlouini:18}. Nevertheless, \ac{FSO} signals are subject to time-variant atmospheric turbulence, which can cause severe degradation in the performance, as the quality of the  \ac{FSO} links continuously varies with time. Therefore, it is of utmost importance to design robust adaptive schemes that can mitigate the aforementioned severe degradation and boost the performance for various channel conditions. 

For  \ac{FSO} systems, the \ac{IM/DD} is preferable over coherent modulation techniques, usually adopted in \ac{RF} based systems, due to its low cost, power consumption, and computational complexity\cite{KhaUysal:14}.  Nevertheless, the signal constellation for \ac{FSO} communications with \ac{IM/DD} adheres to additional constraints compared to the systems operating in the \ac{RF} band. In particular, the input signal is subject to non-negative signaling and average optical power constraints.  The exact capacity of \ac{FSO} channels and the associated capacity-achieving distribution are still open research problems. The authors in \cite{FaridHran:10} derive upper and lower bounds on the capacity of  \ac{IM/DD} optical channels. For this channel, the input is subject to non-negativity and  average optical power constraints.
Upper and lower bounds on the capacity of \ac{IM/DD}  channels are derived in \cite{LapMoser:09}, where the signal is constrained in both its average and peak power.  Upper and lower bounds on the capacity of several optical channels, i.e., \ac{SISO}, parallel, broadcast, and multiple access optical channels are provided in
\cite{ChaMorSlim:16,ChaRezSlim:17,ChaRezkiSlim:17,ChaRezSlim:16,ChaAlEbraNafSlim:17}.

 Efficient \ac{FSO} communications  can be realized by the joint design of higher-order modulation schemes  and channel coding, known in the literature as \ac{CM} \cite{ForGal:84,Ungerboeck:87,GoldsmithChua:98,Djordjevic:May10,ChaLioKar:11,San:11,AnguitaDjordjevic:05}.   
Generally, to design transceivers with a transmission rate close to the channel capacity, three main requirements are considered. In particular, the distribution of the symbols should match the capacity-achieving distribution of the channel, and optimal sufficiently long channel codes are required. Also,  the transmission rate should be adapted with fine granularity according to the channel condition, i.e., the encoder supports a large number of transmission modes over a wide range of \acp{SNR}. Unfortunately, the design of efficient coded modulation systems that fulfill the above requirements is a challenging task. For instance, the input distribution is not, in general, the capacity-achieving one, causing what is called the shaping gap\cite{ForGal:84}. Additionally, finite-length \ac{FEC} codes are implemented in practice, which results in a coding gap\cite{BocSteSch:15}. Finally, the number of allowable modulation orders and coding rates to choose among are limited by the targeted system complexity, leading to partially-adaptive systems. In this work, we focus more on the shaping gap and rate adaptability aspects.

Regarding the shaping gap, optimizing the shape of the modulated signal constellation can decrease the difference between the transmission rate and Shannon's limit. The main categories of constellation shaping are geometric shaping and \ac{PS}. In geometric shaping, the symbols in the constellation are equiprobable and non-uniformly spaced. On the contrary, a probabilistically shaped constellation is uniformly spaced with varying probabilities per symbol. The latter attracted increased attention in the last several years, due to its higher shaping gain,  rate adaptability, and the possibility of using Gray code for symbol labeling \cite{BocSteSch:15,BucSte:15,BocSch:19}.

\subsection{Related Work}
An  \ac{IM/DD} based scheme is proposed for \ac{AWGN} channels with electrical power constraint in \cite{GitMatSte:19}. The coded modulator employed a tailored-designed \ac{LDPC} channel codes with probabilistically shaped \ac{OOK} symbols. The scheme outperforms uniform signaling with around $1$~dB. However, the spectral efficiency is low because of considering  \ac{OOK} rather than higher-order modulations. Therefore, the transmission data rates are limited, i.e., the maximum transmission rate is $1$ \ac{bpcu}. Also, the scheme considers the average electrical power constraint for the signal, which is opposed to the average optical power constraints for intensity channels.  The scheme further assumes an \ac{AWGN} channel, which does not account for the rate adaptability to cope with the diverse channel conditions with turbulence-induced fading in \ac{FSO} channels.

The unipolar \ac{MPAM} signaling is considered as a promising candidate for  \ac{IM/DD} systems, because it can achieve a near-capacity performance for \ac{FSO} channels  \cite{FarHar:09,FaridHran:10}. A lower bound on the capacity  of \ac{FSO} channels  is  proposed with non-uniform \ac{MPAM} signaling in \cite{FarHar:09}.  In particular, the  input distribution is designed to maximize the source entropy, which approximates the optimal distribution at high \acp{SNR}. In fact, the optimal capacity-achieving distribution should maximize mutual information, rather than the source entropy. {\color{black} For the implementation of the scheme in \cite{FarHar:09}, \ac{MLC} with \ac{MSD} is considered. In this scheme, the encoder necessitates \ac{MLC} with multiple \ac{FEC} encoders to generate the probabilistically shaped symbols with the desired distribution. At the receiver,   \ac{MSD} is required, leading to  error propagation and long latency due to the successive decoding of bit levels\cite{ChoWinzer:19}.}\footnote{{\color{black} Note that a parallel architecture for \ac{MSD}, without successive decoding, can reduce the latency and error propagation at the expense of decreasing the achievable rate \cite{ChoWinzer:19}.}}
In \cite{HeBoKim:19}, a sub-optimal coded modulation scheme for \ac{IM/DD} channels is proposed with unipolar \ac{MPAM}. In this approach, only the even-indexed symbols are freely probabilistically shaped, while the probability of the odd-indexed symbols is forced to equal the probability of the preceding even symbol.  In this scheme,  the input distribution can not be fully optimized to match the capacity-achieving distribution of the channel, leading to an increased shaping gap and a rate loss.
To the best of the authors' knowledge,  efficient and practical adaptive coded modulation schemes with fine granularity and their capacity-achieving input distributions have not been well investigated for \ac{IM/DD} \ac{FSO} channels.

In fiber-optical communications,  \ac{PS} has recently gained increased interest, followed by the introduction of the \ac{PAS} scheme to approach the capacity of fiber-optical channels in \cite{BocSteSch:15}.  For the \ac{PAS} architecture, the capacity-achieving distribution should be symmetric around zero. In this case, the uniformly distributed parity bits from the \ac{FEC} encoder can modulate the sign of the symbols. Therefore, it is only suitable for  bipolar input signals \cite{Bocherer:14,YanZibLar:14,BocSteSch:15,BucSte:15,PanKsc:16,SheLiva:18,HuYanDa:18,BocSch:19}.  Unfortunately, the \ac{PAS}   can not be directly extended to \ac{IM/DD} in \ac{FSO} channels,  as the constellation symbols are constrained to be non-negative, i.e., unipolar signaling.
\subsection{Contributions}
In this work, we propose an adaptive coded modulation scheme with fine granularity to approach the capacity of  \ac{IM/DD}  \ac{FSO} channels through \ac{MPAM} signaling with non-negativity and optical power constraints. Realistic models for the atmospheric turbulence  are considered, i.e., Gamma-Gamma and Lognormal distributions \cite{AlhaAndPhi:01,BenRezkiSlim:13}. The proposed encoder considers probabilistic shaping of a unipolar \ac{MPAM} constellation with a  \ac{CCDM}, followed by an efficient \ac{FEC} encoder. Therefore, the information symbols can be probabilistically shaped, while the parity check bits, generated by the \ac{FEC} encoder, are uniformly distributed.  In the decoder, the \ac{FEC} decoding is performed before the distribution dematching (i.e., reverse concatenation architecture). We compute the capacity of the proposed scheme (i.e., the maximum achievable rate) with both optimal \acp{SMD} and practical low complexity \acp{BMD}, for various \acp{SNR}.  The ergodic capacity of the coding scheme is calculated for Gamma-Gamma and Lognormal turbulence fading. The encoder can operate when the \ac{CSI}  is known at the encoder and decoder or only at the decoder. In this regard, we derive the outage probability due to the turbulence-induced fading, when the \ac{CSI} is not available at the transmitter.  The contributions of the manuscript can be summarized as follows.
\begin{itemize}
	\item We propose a coded modulation scheme to achieve the capacity of  \ac{FSO} channels with probabilistically shaped unipolar \ac{MPAM} signaling.
	\item  An algorithm is provided to compute the  input distribution that achieves  the capacity of the coded modulation scheme, utilizing  \acp{SMD} and  \acp{FEC} with an optimized code rate.
	\item For practical channel encoders with a finite set of coding rates and \acp{BMD}, an algorithm is proposed to estimate the distribution which approaches the capacity of the proposed scheme.
	\item The ergodic capacity of the proposed coding scheme is analyzed for different \acp{SNR} and turbulence conditions.
	\item  The outage probability is derived for the blind encoder, i.e., the \ac{CSI} is not known at the transmitter. 
	\item An approach to design the blind encoder with an arbitrary outage probability is proposed.    
\end{itemize}
{\color{black} The proposed transponder has several distinguishing features. First, the \ac{IM/DD}  with \ac{MPAM} signaling has lower computational complexity, and it can be efficiently implemented compared to coherent modulation \cite{KhaUysal:14}.   
Moreover, the architecture considers a \ac{CCDM}, which is asymptotically optimal in the frame length,   and a single binary \ac{FEC} encoder and decoder at the transmitter and receiver, respectively. Also, the rate can be adapted to the \ac{SNR} for various turbulence conditions with fine granularity. Finally, the encoder can operate within a pre-designed maximum outage probability in the absence of the \ac{CSI} at the transmitter.}
\subsection{Paper Organization and Notations}
The work is organized as follows. In \cref{sec.model}, the signal model is described, and the capacity of \ac{MPAM} signals is calculated. In \cref{sec.scheme}, the proposed scheme is introduced,  the capacity of the sparse-dense signaling is obtained for \ac{SMD} and \ac{BMD}. Also,  the optimal operating rate using the proposed encoder is analyzed. In \cref{sec.adaptive}, an algorithm to estimate the capacity of the coding scheme is provided, assuming optimal  \ac{SMD}  with a fully-controllable channel coding rate. {\color{black} For channel encoders with a finite set of possible rates, the capacity and the computational complexity of the  scheme  are analyzed in \cref{sec.adaptivescheme}.}
The outage probability is derived in Section~\ref{sec.outage}, while numerical results are provided in \cref{sec.numericalresults} to attest the scheme performance. Finally, we conclude the work in \cref{sec.conclusion}.

Throughout this paper, we denote \acp{r.v.} with capital letters and their realizations with small letters.  We use $\mathbb{P}\{ X=x \}$ to denote the probability that a discrete \ac{r.v.} $X$ equals $x$. The expression $\mathbb{E}\{X\}$ denotes the expected value of the \ac{r.v.} $X$. For vectors, we use  bold letters, e.g., ${\mathbf x}=[x_{1},x_{2},\cdots, x_{n}]$. All logarithmic functions used throughout the paper are of  base $2$.
%
\section{Signal Model}\label{sec.model}
Let us consider $\n$ transmissions (channel uses) over a discrete-time  \ac{FSO} channel, also known as optical direct-detection channel with Gaussian post-detection noise \cite{HraKac:04,LapMoser:09,FarHar:09}.  In this channel, the input signal modulates  the light intensity, while a photo-detector at the receiver  produces a noisy signal that is proportional to the intensity. The dominant noise sources are  thermal noise,  intensity fluctuation noise by the laser
source, and  shot noise induced by ambient light. The contributions from all the noise sources can be modeled as \ac{AWGN} \cite{LapMoser:09}. Hence, the received signal at time instant $i$ can be written as 
\begin{align}\label{eq.signalmodel}
\Y_{i}=G\,\X_{i}+\W_{i}, &&\text{for } i\in \{1,2,\cdots,n\}, 
\end{align} 
where $\X_{i}$ is the channel input,  $\W_{i}$ is a Gaussian noise with zero mean and variance $\sigma^2$, and $\H$ is a \ac{r.v.} representing the fading due to the atmospheric turbulence. In \ac{FSO} systems, the channel changes slowly with respect to the bit rate. Hence, $\H$ is considered as a block fading process, and it is assumed to be fixed over the entire frame of $n$ symbols. For moderate and strong  turbulence,  the irradiance fluctuations can be modeled as a Gamma-Gamma distribution\cite{AlhaAndPhi:01}. The \ac{PDF} of the Gamma-Gamma \ac{r.v.} is 
\begin{align} \label{eq.pdfgammagamma}
f_{G}\left ({g}\right )=\frac {2\left ({\alpha \beta }\right )^{\frac{\alpha +\beta }{2}}}{\Gamma \left ({\alpha }\right )\Gamma \left ({\beta }\right )}\,g^{{\frac{\alpha +\beta-2 }{2}}} \, K_{\alpha -\beta }\left ({2\sqrt {\alpha \beta g}}\right ),\,\, g>0,
\end{align}
where $K_{a}(\cdot)$ is the modified Bessel function of the second kind of order $a$, and the parameters $\alpha$ and $\beta$ are the effective number of small scale and large scale cells of the scattering environment, which can be expressed as a function of the Rytov variance, $\sigma^{2}_{\text{R}}$\cite{AlhaAndPhi:01}. For example, considering plane waves  from \cite[(14)-(19)]{AlhaAndPhi:01}, we have
\begin{align}
\alpha(\sigma_{\text R})&=\left[\exp \left(\frac{1+0.49 {\sigma}^{2}_{{\text R}}}{\left(1.11 \sigma^{12/5}_{\text R}\right)^{7/6}}\right)-1\right]^{-1},\nonumber \\ 
\beta(\sigma_{\text R})&=\left[\exp \left(\frac{1+0.51  {\sigma}^{2}_{{\text R}}}{\left(0.69 \sigma^{12/5}_{\text R}\right)^{7/6}}\right)-1\right]^{-1}.
\end{align}

A common format for the channel input  $\X_{i}$ that can achieve a near capacity performance for  \ac{IM/DD} systems is unipolar \ac{MPAM}\cite{FarHar:09,FaridHran:10}. Considering that the \acp{r.v.} $\{X_{i}\}_{i=1}^{n}$ are \ac{i.i.d.}, the \ac{PMF} of $X\in \{a_{0},a_{1},\cdots, a_{M-1}\}$ can be written as ${\p \triangleq [p_{0},p_{1},\cdots, p_{M-1}]}$, where  $\aj$ is the $j$th element of $\a \triangleq [0,\Delta,\cdots,(M-1)\Delta]$,   $\Delta >0$ is the spacing between the symbols, and $p_{j}$ is the probability assigned to the constellation symbol $\aj$. Let us define the set that includes all possible symbol distributions as
\begin{align}
\mathbb{S}=\Bigg\{& \p :       \p = [p_{0},p_{1},\cdots, p_{M-1}] , \nonumber \\
&\sum_{j=0}^{M-1} \pj=1,    p_{j}\geq 0, \,\forall j \in \{0,1,\cdots,M-1\}                  \Bigg\}.
\end{align}
The  signal is also subject to an average optical power constraint, i.e., 
\begin{equation}
\mathbb{E}\left\{ {X} \right\} =\sum_{j=0}^{M-1} \pj\, \aj \leq \pw,
\end{equation}
where $\pw$ is the average optical power limit. {\color{black} In fact,  the optical power constraint is more relevant for \ac{FSO} signals  than the  electrical power limit (i.e., $\mathbb{E}\left\{ {X}^{2} \right\} \leq P$), widely adopted in  \ac{RF}  systems\cite{HraKac:04,LapMoser:09}. Also, the instantaneous optical \ac{SNR}, defined as $g \pw/\sigma$, is usually adopted in \ac{FSO} communication, rather than the electrical \ac{SNR}  \cite{HraKac:04,FaridHran:10,LapMoser:09,ChaMorSlim:16}.}

In order to guarantee reliable communication with an arbitrarily low probability of error, the transmission rate should be less than the achievable rate of the coded modulation scheme.\footnote{\color{black} The terms rate and instantaneous rate are interchangeably used  to indicate the  rate at an instantaneous \ac{SNR} for a fixed $g$. On the other hand, the ergodic rate is the average  rate over the irradiance  distribution at a given $\pw/\sigma$.}  The achievable rate depends on the distribution of the signal at the channel input. Hence, the input distribution should be optimized to maximize  the achievable rate.
In this regard, the capacity of unipolar \ac{MPAM}, $C(g)$,   can be found for a given  $g$  as the optimal value of the following optimization problem
\begin{subequations}\label{eq.rpam}
	\begin{alignat}{2}
	&\underset{\Delta>0,\, \p \in \mathbb{S}}{\text{maximize}}\quad && \I(X;Y|G=g) \\
	&\text{subject to} &&\a^{T}\,\p \leq  P 
	\end{alignat}
\end{subequations}
where  $\!\I(X;Y|G\!=\!g) \!=\! {h}(Y|g) \!-\! {h}(Y|X,g)$ is the mutual information between $X$ and $Y$ given $\!G\!=\!g$, 
\begin{align}\label{eq:diffentropy}
&{h}(Y|g) \triangleq -{ \int}_{-\infty}^{\infty} P_{{Y}|G}(y|g) \log P_{{Y}|G}(y|g) \, \d y, \\
& {h}(Y|X,g)=h(\W)= \log\left( \sqrt{2\, \pi\, e\, }\sigma\right) \label{eq:conddiffentropy}
\end{align}
are differential  conditional  entropy functions, and
\begin{equation}\label{pdfy}
P_{Y|G}(y|g)= \frac{1}{\sqrt{2 \pi \sigma^2}}\, {\sum_{j=0}^{M-1}}\, p_{j}\,  \exp\left(- \frac{{\left(y-g\,\aj\right)}^{2}}{2\, \sigma^2}\right)
\end{equation}
is the distribution of the received signal conditioned on the channel gain.\footnote{In this work, we refer to the maximum achievable rate of a specific scheme by the capacity of that scheme.}
For notation simplicity, let   ${\Ip \triangleq \I(X;Y|G=g)}$ to emphasize its dependence on the parameters of the input distribution, i.e., $\p$ and $\Delta$.
{\color{black} For a fixed $\Delta$, the {\color{black} problem} \eqref{eq.rpam} is a convex optimization problem in $\p$ \cite[Theorem 2.7.4]{CoverThomas:06}. In fact, $\Ip$ is a concave function in $\p$ from the concavity of the conditional entropy function in \eqref{eq:diffentropy} and   the composition with an affine mapping property \cite[Sec. 3.2.2]{BoydVan:04}}. Hence, it can be efficiently solved using any suitable convex optimization algorithm, e.g., interior-point method, and the optimal $\p$ at this $\Delta$, $\p_{\Delta}$, can be computed. Now, it is required to find the optimal value of the constellation spacing that maximizes the mutual information. 
Since ${\mathbb I}_{\Delta}(\p_{\Delta}|g)$ is unimodal in $\Delta$, any efficient optimization algorithm in one dimension can be employed to compute  the optimal constellation spacing, $\Delta^{*}$, e.g.,  golden section search and successive parabolic interpolation\cite{BocSteSch:15}. 

In order to achieve the maximum capacity, the channel input $X$ should be probabilistically shaped to have a distribution with the optimal parameters obtained from \eqref{eq.rpam}. Unfortunately, there is no known computationally efficient and practical scheme in the literature which achieves this rate for unipolar \ac{MPAM} signals. For example, the efficient \ac{PAS} encoder, usually adopted for coherent fiber optical communications, can not be used due to the asymmetric signaling around zero in unipolar \ac{MPAM} \cite{BocSteSch:15}. 
\begin{figure*}
	\centering
	\includegraphics[width=0.992\linewidth]{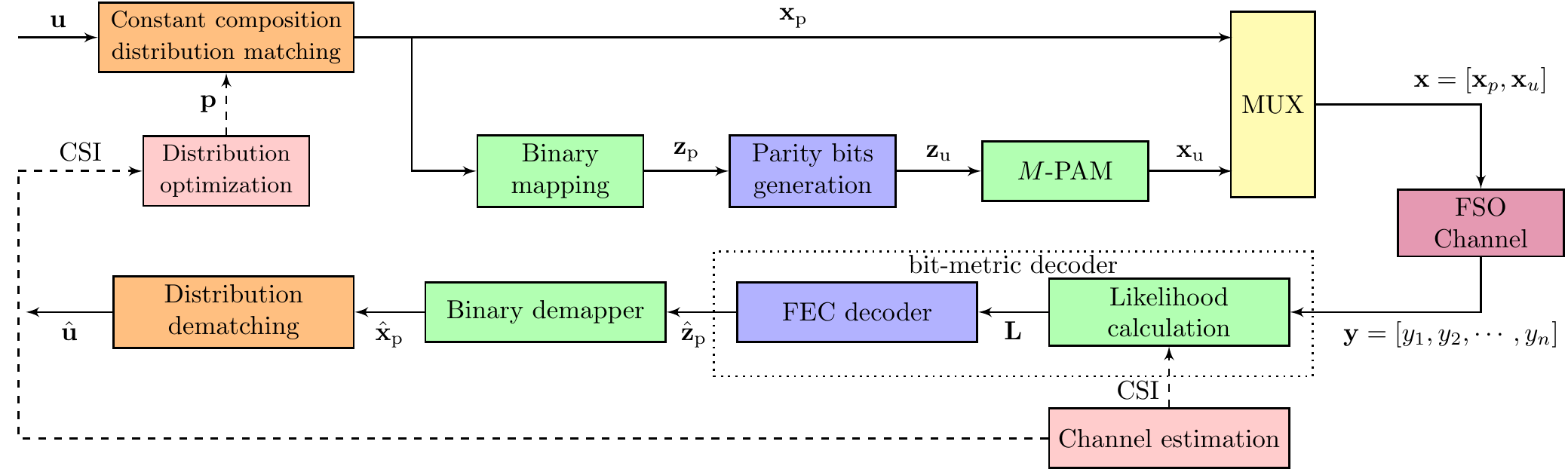}
	\caption{The proposed  probabilistic shaping  scheme using unipolar \ac{MPAM} for \ac{FSO} communications.}
	\label{fig:encodercsi}
\end{figure*}
\section{Proposed Coded Modulation scheme with \acs{MPAM} for \acs{FSO}}\label{sec.scheme}
We propose two practical adaptive coded modulation schemes to increase the bandwidth efficiency of \ac{FSO} communications by probabilistically shaping the input distribution. The first method considers that the \ac{CSI} is available at both the encoder and decoder, as shown in  \cref{fig:encodercsi}. For the second scheme, the \ac{CSI} is assumed to be known only at the receiver (blind encoder). In this section, we start by describing the encoder, and the maximum mutual information between the unipolar $\ac{MPAM}$ signal at the channel input and the received noisy signal. Then, the decoder, its achievable rate, and the optimal feasible operating rate are discussed. 
\subsection{Sparse-Dense  Encoder}
The target of the encoder is to convert the input binary string of uniformly distributed bits into probabilistically shaped unipolar \ac{MPAM} symbols and to perform channel coding. Hence, the decoder can reliably recover the original data from noisy measurements at the receiver. The proposed scheme is described in detail below. 

\subsubsection{Capacity-achieving Distribution}
First, we compute the capacity-achieving distribution of the proposed scheme, parameterized with $\p^{*}$ and $\Delta^{*}$, which permits the highest reliable communication rate for a given \ac{SNR}. Section~\ref{sec.adaptivescheme} proposes an algorithm to obtain the optimal distribution when  \ac{CSI} is known at the encoder, or when only channel statistics are available, as in Section~\ref{sec.outage}.

\subsubsection{Distribution Matching} \label{Sec.CCDM}
The distribution matching transforms the uniformly distributed input bit string, ${\mathbf u}\in \{0,1\}^{k_{\text p}}$,  into unipolar \ac{MPAM} symbols,  ${{\mathbf x}_{\text p} \in \{0,\Delta^{*},\cdots, (M-1)\Delta^{*} \}^{n_{\text p}}}$ with the target distribution, $\p^{*}$. Several \ac{DM} techniques  have been proposed in the literature with various computational complexity, rate loss, and  parallelization ability\cite{SchBoc:16,BocSteSch:17,PikWenKramer:19,FehMilKoiPar:19,FehMilKoiPars:20}. The \ac{CCDM} is an invertible mapping from a fixed-length vector of uniformly distributed bits  to a fixed-length sequence of shaped symbols (i.e., amplitudes) \cite{SchBoc:16}. 
	The empirical distributions of all possible output sequences are identical, i.e., they have a constant composition. Therefore, every output sequence follows, to some extent, the target distribution.  The target distribution should be quantized such that the probability of each symbol can be represented as a rational number, where the denominator is the frame length length, $n_{\text p}  $. In other words,   the \ac{PMF} of $X_{\text p}$, $\p^{*}$, is approximated by what is called $n_\text{p}$-type distribution in the form of  $\tilde{\p}\triangleq \left[z_{0}/ n_{\text p},z_{1}/ n_{\text p}, \cdots, z_{M-1}/ n_{\text p}   \right]$, where $z_{j}$ is an integer representing the number of times  at which the  $j$th symbol appears and  $\sum_{j=0}^{M-1} z_{j}= n_{\text p}$.  The discrepancy between the target and $n_\text{p}$-type distributions   decreases with the output sequence length. Hence, for asymptotically large $n_{\text p}$, the quantization error for $\p^{*}$ is negligible, i.e., $\lim_{n_{\text p}\rightarrow \infty} \tilde{\p} = \p^{*}$.
	
	In order to quantify the \ac{CCDM} rate,  $R_{\text{DM}}\triangleq  {k_{\text p}  }/{ n_{\text p}}$, the number of input bits should be computed. The number of bits, $  k_{\text p}  $, that are required to be transformed to $n_{\text p}$ shaped symbols depends on  the number of possible configurations (i.e., permutations) of the output symbols that have empirical distribution $\tilde{\p}$. More precisely, we have
	\begin{align} 
	{k}_{\text p}&= \Bigg\lfloor\log \left(\begin{array}{c} {n}_{\text p}\\
	{z_0,z_1,\ldots,z_{M-1}} 
	\end{array}\right) \Bigg\rfloor \nonumber \\ 
	&=\Bigg\lfloor \log \left( \frac{{n}_{\text p}\!}{\prod_{j=0}^{M-1}\,  \left({z}_{j} !\right)\,}\right) \Bigg\rfloor,
	\end{align}
	where ${z}_{j} !$ is the factorial of ${z}_{j}$ and $\begin{tiny}
	\left(\begin{array}{c} \cdot\\
	{\cdot,\cdot,\cdot}
	\end{array}
	\right)\end{tiny} $ is the multinomial coefficient that determines the number of permutations \cite[24.1.2 ]{AbrSte:72}.
	The rate of the  \ac{CCDM}, i.e., the number of bits per output symbol, converges to the entropy of the source, for asymptotically large number of output symbols\cite{SchBoc:16}, i.e.,
	\begin{equation}\label{eq:dmrate}
	\lim\limits_{n_{\text p} \rightarrow \infty}\,  \frac{  k_{\text p}  }{ n_{\text p}  } = \Hp,
	\end{equation}
	where $\Hp \triangleq -\sum_{j=0}^{M-1} \pj\,\log(\pj) $ is the entropy of the discrete \ac{r.v.}. 
	%
	On the other hand, for finite block lengths, the \ac{CCDM} exhibits a rate lower than the source entropy. The rate loss can be upper bounded from \cite{SchBoc:16}  as
	\begin{align}
	R_{\text{loss}}&\triangleq \mathbb{H}(X_{\tilde{\text{p}}})-R_{\text{DM}} \nonumber \\
	 &\leq \frac{1+ (M-1) \log(n_{\text p}+M-1) }{n_{\text p} },
	\end{align}
	where $\mathbb{H}(X_{\tilde{\text{p}}})$ is a entropy of the $n_\text{p}$-type distributed \ac{r.v.} $X_{\tilde{\text{p}}}$.
	For example, ${R_{\text{loss}}< 7.5 \times 10^{-4}}$ bits/symbol for $8$-\acs{PAM} with  frame length $n_{\text p}=64800$, adopted in the \ac{DVB-S2} \cite{DVB:14}.
	
	For the invertible mapping between the bits and the symbols, large \ac{LUT} can be used, in principle. However, for long block lengths, the size of the table is too large to be useful, i.e., $2^{R_{\text{DM}}\, n_{\text p}}$. Alternatively,  the mapping can be achieved in an algorithmic manner, e.g., arithmetic coding is considered for the \ac{CCDM} implementation in \cite{SchBoc:16}.  
\subsubsection{Channel Coding}
In order to achieve reliable communication with high spectral efficiency close to the channel capacity, an \ac{FEC} scheme should be employed. Since one of the main targets is to keep the computational complexity low,  we opt for binary \ac{FEC} encoders, as they have low complexity compared to non-binary methods.  In this regard, the probabilistically shaped \ac{MPAM} signal is first mapped into a binary string using a  mapper ${\mathbb B}$, where each element of ${\mathbf x}_{\text p}$ is labeled by $m\triangleq \log(M)$ bits, i.e,
{\color{black} 
\begin{align}
{ { { \mathbb B}}}\left(x_{{\text p}_i}\right)= \left[b_{{i,1}},b_{{i,2}},\cdots, b_{{i,m}}\right],\,\,\text{for } i\in \{1,2,\cdots,n_{\text p}\},
\end{align}
where $b_{i,\ell}$ is the $\ell$th bit level of the $i$th symbol.
The vector ${\color{black} {\mathbf b}_{\ell}\triangleq\left[ b_{{1,\ell}},b_{{2,\ell}},\cdots, b_{{n_{\text p},\ell}}\right]}$ contains all the bits of level $\ell \in \{1,2,\cdots,m\}$.}
A single binary string, ${\mathbf z}_{\text p} \in {\{0,1\}}^{m\, n_{\text p}}$, is formed from the concatenation of the mapped bits for all the $n_{\text p}$ symbols, where  
\begin{equation}
{\mathbf z}_{\text p} \triangleq \left[{\mathbb B}\left(x_{{\text p}_1}\right),{\mathbb B}\left(x_{{\text p}_2}\right),\cdots,{\mathbb B}\left(x_{{\text p}_{n_{\text p}}}\right)\right].
\end{equation}
The proper choice of the binary mapper ${\mathbb B}$ improves the performance of the scheme, e.g., the reflected binary mapping (Gray code) yields good performance \cite{BocSteSch:15}.

In this scheme, any systematic binary \ac{FEC} encoder with rate $\rc$, dimension $\tilde{k} \triangleq m\, n_{\text p}$, and block length $\tilde{n} \triangleq \tilde{k}/\rc$ can be employed. 
The redundant information in terms of the parity bits can be generated by 
\begin{equation}
{{\mathbf z}_{\text u}}={\mathbf P}^{T} {{\mathbf z}_{\text p}},
\end{equation}
where the multiplication is in the Galois field of two elements, and $ {\mathbf P}\in \{0,1\}^{\tilde{k}\times (1-\rc)\tilde{n}}$ can be found by putting the code generator matrix   in the standard form, ${\begin{bmatrix}\mathbf{I}_{\tilde{k}}|\mathbf{P}\end{bmatrix}}$, with $\mathbf{I}_{\tilde{k}}$ denoting the  $ \tilde{k}\times \tilde{k}$ identity matrix.

Although the vector ${{\mathbf z}_{\text p}} $ at the input of the \ac{FEC} is probabilistically shaped, the parity bits, ${\mathbf z}_{\text u}$, tend to be uniformly distributed. This is attributed to the fact that each redundancy bit results from a modulo-$2$ sum of a large number of bits \cite{Mackay:99,BocSteSch:15}. 
{\color{black} These parity bits are mapped to the corresponding unipolar \ac{MPAM} symbols by binary demapper $\mathbb{B}^{-1}$, i.e.,
	\begin{eqnarray}
	\mathbb{B}^{-1}({{\mathbf z}_{\text u}} )={{\mathbf x}_{\text u} \in \{0,\Delta^{*},\cdots, (M-1)\Delta^{*} \}^{(1-\rc)\,n}},
	\end{eqnarray}
	 which are also uniformly distributed. }
\subsubsection{Sparse-Dense Transmission}
{\color{black} The parity symbols, ${\mathbf x}_{\text u}$, are appended to the probabilistically shaped symbols, ${\mathbf x}_{\text p}$, to form the codeword ${\mathbf x}=\left[{\mathbf x}_{\text p},{\mathbf x}_{\text u} \right]$, with $n= {n}_{\text p}/\rc=\tilde{n}/m$ symbols.} Since part of the time (channel uses) is dedicated to the shaped symbols while the other part is reserved for uniform symbols, this scheme can be considered as a time-sharing encoder. From another perspective, if one considers the amount of information in each part, the system can be regarded as a \acf{SDT} scheme. The reason behind this is that the uniform distribution maximizes the source entropy (dense information representation), while the probabilistically shaped symbols have less amount of information (sparse). In the following, we refer to the proposed scheme by \ac{coms}.  
\subsection{Capacity of the Sparse-Dense Transmission}\label{sec:achratetx}
The capacity of sparse-dense signaling can be considered as an upper bound on the achievable rate of our system, for a given \ac{FEC} rate.  Hence, it is beneficial to compute the maximum mutual information (i.e., capacity) of the  \ac{SDT}, regardless if such a rate is achievable or not by the proposed  \ac{coms} scheme. First, let us define $X_\text p$ as the \ac{r.v.} representing the probabilistically shaped symbols, $\{X_{i}\}_{i=1}^{n\,{\rc}} $, and $X_\text u$ as  the \ac{r.v.} representing the uniformly distributed symbols, $\{X_{i}\}_{i=n\,{\rc}+1}^{n}$.  The capacity of the \ac{SDT}, $\cts(g)$, can be found from  \eqref{eq.signalmodel} and \eqref{eq.rpam} as the optimal value for the following optimization {\color{black} problem}
\begin{subequations}\label{eq.optts}
\begin{alignat}{2}
&\!\!\!\!\underset{\Delta>0,\p\in \mathbb{S}}{\text{maximize }}
&&\rts(g,\Delta,{\mathbf p})\triangleq \rc \,\Ip+(1\!-\!\rc)\Iu\\
&\!\!\!\! \text{subject to}
& & \rc  \,\a^{T} \p +0.5\, \Delta {(1- \rc)(M-1)}\,\leq {P}, \label{eq.powerconstpu}
\end{alignat}
\end{subequations}
where {\color{black} $p_{j}\triangleq \mathbb{P}\{ X_\text p=a_{j}\}$ and $u_{j} \triangleq \mathbb{P}\{ X_{\text u}=a_{j} \}=1/M$, for $j\in \{0,1,\cdots,M-1 \}$.
{\color{black} Similar to \eqref{eq.rpam}, the optimization problem is convex in the probability vector ${\p}$,  for a fixed $\Delta$, where $(1-\rc)\Iu$ and $0.5\, \Delta {(1- \rc)(M-1)}$ are constants that do not depend on $\p$ \cite{CoverThomas:06}}. Hence, the interior-point algorithm can be used to find the optimal probabilities. Regarding  $\Delta$, the golden section method can be adopted to obtain the value of $\Delta$ that maximizes the achievable rate within its feasibility range. The range of $\Delta$ that satisfies the power constraint can be found from \eqref{eq.powerconstpu} as
\begin{align}
\Delta &\leq  \frac{P}{\rc  \,\sum_{j=0}^{M-1} j\, p_{j} +0.5\,  {(1- \rc)(M-1)}} \nonumber\\
& \leq \frac{2 \pw}{(1-\rc)(M-1)},
\end{align}
where the second inequality holds with equality if ${\p=[1,0,\cdots,0]}.$}
The \ac{SDT}  capacity can now be expressed, for a given $g$,  as $\cts(g)=\rts(g,\Delta^{*},{\mathbf p}^{*})$, where $\Delta^{*}$ and ${\mathbf p}^{*}$ are the optimal symbol spacing and probabilities obtained as the solution of optimization problem \eqref{eq.optts}.  
\subsection{Bit Metric Decoder}\label{sec.BMD}
In the decoder, the \ac{FEC} decoding is performed before the distribution dematching (i.e., reverse concatenation architecture). The reverse concatenation method  prevents the common problem of the burst of errors after the distribution dematching, due to the receipt of erroneous symbols from the channel \cite{ChoWinzer:19}. For the \ac{FEC} decoding, the capacity of the \ac{SDT}, computed in the \cref{sec:achratetx}, could be achieved  using an optimal \acf{SMD}. However,  the computational complexity of \ac{SMD} is high. On the contrary, a \acf{BMD} can yield a rate close to  $\cts(g)$, while having lower complexity. For \ac{BMD} with soft decisions,  one real number is computed for each bit level of each received symbol,  representing the likelihood of this bit. The \ac{LLR} of the $\ell$th bit level can be written given  the received symbol $y_{i}$ as
 \begin{align}
 L_{i,\ell}&=\log \frac{f_{B_{i,\ell}|Y_{i},G}(0|y_{i},g)          }{f_{B_{i,\ell}|Y_{i},G}(1|y_{i},g)}
\nonumber \\&=\log \frac{\sum_{x\in \mathcal{X}_{\ell}^{0}} f_{Y_{i}|X,G}(y_{i}|x,g) \mathbb{P}\{ X_i=x \}          }{\sum_{x\in \mathcal{X}_{\ell}^{1}} f_{Y_{i}|X,G}(y_{i}|x,g)\mathbb{P}\{ X_i=x \}},
 \end{align} 
{\color{black} where $g$ is the value of the channel gain which is correctly estimated at the receiver,  $B_{i,\ell}$ is a \ac{r.v.} representing the $\ell$th level bit of the $i$th symbol, and the sets $\mathcal{X}_{\ell}^{0}$ and $\mathcal{X}_{\ell}^{1}$ include all the values of $x$ such that the $\ell$th level of their binary mapping equals $0$ and $1$, respectively.}
Since the distribution of the signal $X_{i}$ depends on the time instant $i$ for the  \ac{SDT} scheme, the \ac{LLR} can be written as
 \begin{equation} \label{eq.LLR}
 L_{i,\ell}=
 \begin{dcases}
 \log \frac{\sum_{x\in \mathcal{X}_{\ell}^{0}} f_{Y_{i}|X,G}(y_{i}|x,g) \mathbb{P}\{ X_{\text p}=x \}          }{\sum_{x\in \mathcal{X}_{\ell}^{1}} f_{Y_{i}|X,G}(y_{i}|x,g) \mathbb{P}\{ X_{\text p}=x \}   },\\
 \qquad \text{for }i\in\{1,2,\cdots,\rc\,n\}\\
 \log \frac{\sum_{x\in \mathcal{X}_{\ell}^{0}} f_{Y_{i}|X,G}(y_{i}|x,g)        }{\sum_{x\in \mathcal{X}_{\ell}^{1}} f_{Y_{i}|X,G}(y_{i}|x,g)  },\\
     \qquad \text{for } i\in\{\rc\,n+1,\rc\,n+2,\cdots,n\}.
 \end{dcases}
 \end{equation} 
Since the \acp{LLR} are the sufficient statistics to recover the transmitted bits, they are provided to the \ac{FEC} decoder for soft-decision decoding of the binary bits. Finally, the estimated bits are mapped into the corresponding symbols. The  distribution dematching maps the first $n_{\text p}$ estimated symbols into their $k_{\text p}$ associated bits.   
 
 The achievable rate of the \ac{BMD} for \ac{PAS} has been investigated in \cite{MarFab:09,BocSteSch:15}. For \ac{SDT}, let us first define the \acp{r.v.}   ${\mathbf B}_{\text p}=\left[B_{{\text p}_{1}}, B_{{\text p}_{2}},\cdots, B_{{\text p}_{m}}\right] \triangleq {\mathbb B}\left(X_{\text p}\right)$, ${\mathbf B}_{\text u} \triangleq {\mathbb B}\left(X_{\text u}\right)$, $Y_{\text p}\triangleq X_{\text p}+W$, and $Y_{\text u}\triangleq X_{\text u}+W$.  
 Then, an  achievable rate with \ac{BMD} for the \ac{SDT} can be written as
 \begin{align}\label{rateBMD}
 R_{\text{BMD}}(g,\Delta,\p)
 & =(1-\rc) \left[{\mathbb H}\left({\mathbf B}_{\text u } \right) - \sum_{\ell=1}^{m}  {\mathbb H}\left({B_{{\text u}_{\ell}}|Y_{\text u},G} \right)  \right]^{+} \nonumber \\
 &\phantom{=}+ \rc \left[{\mathbb H}\left({\mathbf B}_{\text p}\right) - \sum_{\ell=1}^{m}  {\mathbb H}\left({ B_{{\text p}_{\ell}}|Y_{\text p},G} \right)\right]^{+} \nonumber  \\
 &= (1-\rc) \left[m- \sum_{\ell=1}^{m}   {\mathbb H}\left({B_{{\text u}_{\ell}}|Y_{\text u},G} \right)\right]^{+}  
\nonumber \\
 &\phantom{=}+ \rc \left[\Hp - \sum_{\ell=1}^{m}  {\mathbb H}\left({B_{{\text p}_{\ell}}|Y_{\text p},G} \right)\right]^{+}  ,  
 \end{align}   
 where $[x]^{+}\triangleq \max(x,0)$ gives the maximum between $x$ and zero. Equation \eqref{rateBMD} is due to the one-to-one mapping between $X$ and its binary vector representation ${\mathbb B}(X)$.  The maximum achievable rate using \ac{BMD} can be found by maximizing $R_{\text{BMD}}(g,\Delta,\p)$ subject to an average power constraint. Nevertheless, the problem is not convex \cite{Boch:14}; hence, we propose to obtain an achievable rate (not the maximum) by considering the distribution obtained by solving  \eqref{eq.optts}. The achievable rate of the \ac{SDT} with \ac{BMD} can now be written as  $C_{\text{BMD}}(g) \triangleq R_{\text{BMD}}(g,\Delta^{*},\p^{*})$.\footnote{Note that $C_{\text{BMD}}(g)$ represents an achievable rate for the \ac{BMD} and not its capacity.} Although the achievable rate of \ac{BMD} is less than that  provided by \ac{SMD}, the rate loss is small, as illustrated in the numerical results.

 Another metric to quantify an achievable information rate is the \ac{GMI}, defined in \cite[Eq. (12)]{ChoWinzer:19}. The \ac{GMI} quantifies the number of transmitted bits per symbol in a way similar to what mutual information does \cite{AlvAgrLavMah:15,ChoSchWin:17,AlvFehCheWil:18}. However, the \ac{GMI} considers a mismatched decoding metric, e.g., \ac{BMD}, in contract to the implied optimal decoder  to achieve the rate indicated  by the mutual information. 

   For \ac{PAS}, it has been shown  that the  \ac{GMI}   equals the achievable rate with \ac{BMD}\cite{ChoWinzer:19}.
   \footnote{The definition of the \ac{GMI} in \cite{ChoWinzer:19} does not account for the optimization over various decoding metrics with the same codeword ranking performance.} 
   For the proposed \ac{PS} scheme with \ac{SDT}, the \ac{GMI} has the same expression as the achievable rate of the proposed scheme under \ac{BMD}, $R_{\text{BMD}}(g,\Delta,\p)$. This can be proved starting from \cite[Eq. (13)]{ChoWinzer:19} and by noting that the transmitted bits are probabilistically shaped only for $100 \,\rc\,\%$ of the channel uses.

In order to achieve reliable communication, the transmission rate should be less than the maximum achievable rate for the proposed \ac{coms} with  \ac{BMD}. Hence, it is essential to determine the transmission rate of the \ac{coms}. Since the information bits need to be transmitted are $k_{p}$ bits in $n$ channel uses, the overall transmission rate can be written from \eqref{eq:dmrate} when  $n\rightarrow \infty$ as
\begin{equation}\label{eq.txrate}
R(\p)=\frac{{k_{\text p}}}{n}=\rc \frac{{k_{\text p}}}{n_{\text p}}  =\rc\, \Hp.
\end{equation}
It is clear from~\eqref{eq.txrate} that the transmission rate depends on the input distribution and the \ac{FEC} rate. Therefore, it is essential in the system design to quantify the  maximum \ac{FEC} rate such that the transmission rate is achievable, i.e., $R(\p)  \leq R_{\text{BMD}}(g,\Delta,\p)$. In the literature, the achievable binary code rate and \ac{NGMI} metrics are usually adopted to quantify the number of information bits per transmitted bits\cite{AlvAgrLavMah:15,ChoSchWin:17,ChoWinzer:19,AlvFehCheWil:18}. In our scheme, the achievable binary code rate is represented from \eqref{rateBMD} and \eqref{eq.txrate}  for a given $\p$ as
\begin{equation}
\mathrm{ABR} = \frac{\left[m- \sum_{\ell=1}^{m}   {\mathbb H}\left({B_{{\text u}_{\ell}}|Y_{\text u},G} \right)\right]^{+}}{\Hp- \left[\Hp - \sum_{\ell=1}^{m}  {\mathbb H}\left({B_{{\text p}_{\ell}}|Y_{\text p},G} \right)\right]^{+} +\chi},
\end{equation}
where $\chi \triangleq\left[m- \sum_{\ell=1}^{m}   {\mathbb H}\left({B_{{\text u}_{\ell}}|Y_{\text u},G} \right)\right]^{+}$. The \ac{NGMI} has the same expression as the achievable binary code rate for \ac{PAS}, under some conditions on the decoding metric, as illustrated in \cite{YosAl:19}.   
\subsection{Optimal Operating Point}\label{sec:operatingpoint}
\begin{figure*}[t!]
	\centering
	\begin{subfigure}{0.5\textwidth}
		\centering
		\includegraphics[width=0.992\linewidth,clip]{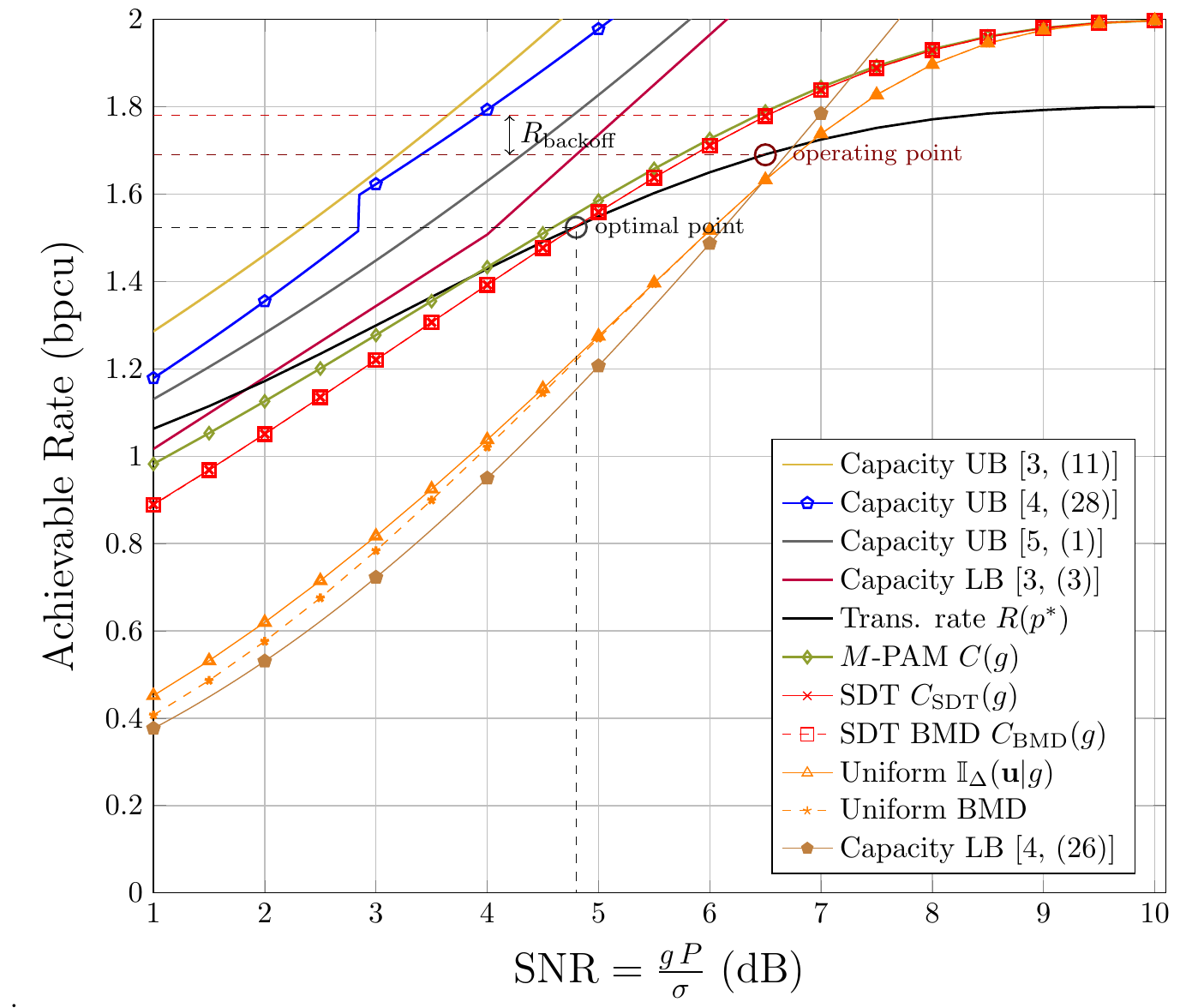}
		
		\caption{}
		\label{fig:m4achievablea}
	\end{subfigure}%
	~
	\begin{subfigure}{0.5\textwidth}
		\centering
		\includegraphics[width=0.992\linewidth,clip]{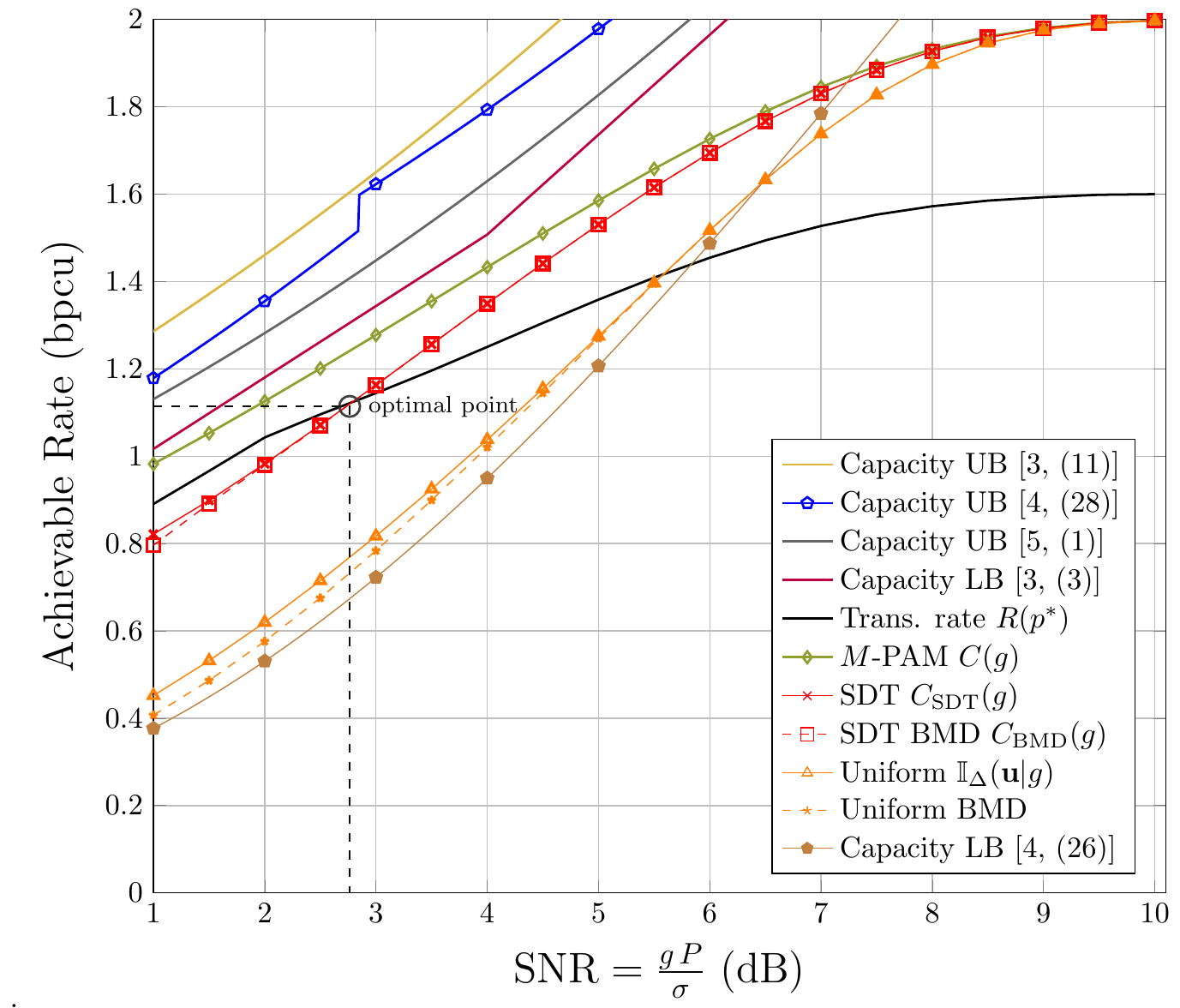}
		\caption{}
		\label{fig:m4achievableb}
	\end{subfigure}
	\caption{The achievable rate vs  \ac{SNR} of various schemes for $M=4$, along with the upper bounds on the capacity of  intensity channels \cite[(11)]{FaridHran:10}, 	\cite[(28)]{LapMoser:09}, and	\cite[(1)]{ChaMorSlim:16}  and the lower bounds in 	 \cite[(3)]{FaridHran:10} and \cite[(26)]{LapMoser:09}:  (a)~$\rc=0.9$, and (b)~$\rc=0.8$.} 
	\label{fig:m4achievable}
\end{figure*}%
For a fixed \ac{FEC}, the optimal distribution $\p^{*}$ that maximizes the achievable rate in \eqref{eq.optts} can lead to an unachievable transmission rate, i.e., $R(\p^{*}) >R_{\text{BMD}}(g,\Delta^{*},\p^{*}) $. 
In this regard, we investigate the optimal operating \ac{SNR} that permits the maximum transmission rate such that reliable communication is still feasible.
First, we describe the case when the \ac{CSI} is known at both the transmitter and receiver.  For a given instantaneous optical \ac{SNR}, $g \pw/\sigma$, the optimal distribution that maximizes \eqref{eq.optts} is calculated.  Accordingly, the capacity $\cts(g)$, the achievable rate with \ac{BMD}, and the transmission rate are computed from \eqref{eq.optts}, \eqref{rateBMD}, and \eqref{eq.txrate}, respectively. 
Then, the optimal operating point is the intersection between the achievable and  transmission rates, i.e., $R(\p^{*})=R_{\text{BMD}}(g,\Delta^{*},\p^{*})$. In order to achieve reliable communication at the optimal operating point, an asymptotically long \ac{FEC} code should be used. For practical finite-rate codes,  lower rates should be considered, e.g., $R=R_{\text{BMD}}-R_{\text{backoff}}$, where  $R_{\text{backoff}}>0$ is a back-off rate to account for the non-optimal \ac{FEC} codes. 

In Fig.~\ref{fig:m4achievable}, the capacity,  achievable rate, and transmission rate  of \ac{BMD}, \ac{SDT}, and uniform signaling  are depicted for various \acp{SNR}. Several  upper and lower bounds on the capacity of intensity channels, i.e.,  \cite[(3) and (11)]{FaridHran:10}, 	\cite[(26) and (28)]{LapMoser:09}, and	\cite[(1)]{ChaMorSlim:16},  are shown in Fig.~\ref{fig:m4achievable}. For $c=0.9$ in Fig.~\ref{fig:m4achievablea}, it can be seen that the optimal \ac{SNR} is  $4.8$~dB, leading to a transmission rate of $1.524$~\ac{bpcu}. The rate gap with respect to the tightest lower and upper bounds on the capacity of \ac{IM/DD}  channel, i.e., \cite[(3)]{FaridHran:10}  and  \cite[(1)]{ChaMorSlim:16}, is $0.15$ and $0.25$, respectively. An example of a practical operating point for finite-length \ac{FEC} codes is provided, which follows the transmission rate curve.  We can see that for  \acp{SNR} lower than the optimal point, the transmission rate is unachievable, while at higher \acp{SNR}, the rate gap increases.

For Fig.~\ref{fig:m4achievableb}  with $\rc=0.8$, the optimal operating point is at  \ac{SNR} of  $2.8$~dB and transmission rate of $1.115$~\ac{bpcu}, which are lower than their corresponding values for $\rc=0.9$.  Also, we can notice that the difference between the achievable rate of \ac{MPAM}, i.e., all the symbols can be probabilistically shaped,  and  
\ac{SDT} is $0.125$~\ac{bpcu} for $c=0.8$, which is larger than the corresponding value for $\rc=0.9$, i.e., $0.075$~\ac{bpcu}. The reason is that the  number of uniformly distributed symbols (not probabilistically shaped) is inversely proportional to the coding rate. 

Regarding the blind encoder, i.e., the \ac{CSI} is available only at the decoder, we can not guarantee that the transmission rate is always less than the achievable rate. This is attributed to the fact that the achievable rate is a \ac{r.v.}, as it is a function of the channel irradiance fluctuation. Hence, there is a non-zero probability that the transmission rate is not achievable, leading to an outage probability. In this case, the encoder can be designed with the worst-case channel condition to guarantee that the outage probability is less than a predefined threshold. This can be achieved by considering a fixed channel gain $g=\bar{g}$ such that the outage probability is upper bounded by the target level, as described in Section~\ref{sec.outage}.

\section{The capacity of  \acs{coms} with \ac{SDT} and Optimal \ac{FEC} Rate}\label{sec.adaptive}
In the previous section, we obtain a single optimal point for each \ac{FEC} rate, i.e., the rate at which the  \ac{SDT} capacity equals the transmission rate. {\color{black} In this section, we provide a technique to obtain the maximum achievable transmission rate of the proposed scheme and the corresponding input distribution for any given \ac{SNR}. This can be done by also optimizing the rate of the channel encoder. Additionally, in order to guarantee that the transmission rate is achievable for the considered  \ac{SNR}, an additional constraint is added such that the input distribution yields an achievable transmission rate.}
More precisely,  the capacity of the proposed scheme can be formulated as
\begin{equation}\label{eq.optts2}
\begin{aligned}
&\underset{\Delta>0,\, \p\in \mathbb{S},\,0<c \leq 1}{\text{maximize}}
& &\rts(g,\Delta,{\mathbf p})\\
& \text{subject to}
& &
\rc\,\a^{T} \p + (1- \rc)\, \Delta\,\frac{M-1}{2}\,  \leq {P},\\
&&&R(\p)= \rts(g,\Delta,{\mathbf p}).\\
\end{aligned}
\end{equation}
%
In order to simplify \eqref{eq.optts2}, the rate constraint can be eliminated by choosing the  channel coding rate such that the achievable rate equals the transmission rate. In particular, the \ac{FEC} rate is written for a fixed ${\mathbf p}$  from \eqref{eq.optts} and \eqref{eq.txrate} as 
\begin{equation}\label{optc}
\rc({\mathbf p})=\frac{\Iu}{{\mathbb H}(X_{\text p})-\Ip+\Iu}.
\end{equation}
Since the transmission rate equals the achievable rate, it is equivalent to maximize $R(\p)$ or $\rts(g,\Delta,{\mathbf p})$. Then, by substituting the channel coding rate into the \eqref{eq.optts2}, the optimization problem becomes
\begin{equation}\label{eq.optts3}
\begin{aligned}
& \underset{\Delta>0,\, \p\in \mathbb{S}}{\text{maximize}}
& &\frac{\Hp\,\Iu}{{\mathbb H}(X_{\text p})-\Ip+\Iu}\\
& \text{subject to}
& &\beta_{\Delta} \bigg[\Hp-\Ip\bigg]
 +\a^{T}\p -{P} \leq 0,
\end{aligned}
\end{equation}
where $\beta_{\Delta} \triangleq (0.5\Delta\,({M-1})- {P})/\Iu$ and the equality constraint is eliminated, as it is always active because of the proper formulation of the coding rate in \eqref{optc}.
Let us define the optimal value of the maximization problem by $\ror(g)\triangleq \rts(g,\Delta,\p^{*})$, where the \ac{PMF}  $\p^{*}$ maximizes \eqref{eq.optts3}  for a given $\Delta$. Since  $\ror(g)$ is the maximum possible rate after the \ac{PMF} optimization, we have
\begin{align}\label{eq.ratediff}
\frac{\Hp\,\Iu}{{\mathbb H}(X_{\text p})-\Ip+\Iu} \leq \ror(g) 
\end{align}
with equality if $\p=\p^{*}$. By multiplying both sides by ${\mathbb H}(X_{\text p})-\Ip+\Iu \geq 0$, equation \eqref{eq.ratediff} can be written as
\begin{multline}\label{eq.ratediff2}
\ror(g) \left[{{\mathbb H}(X_{\text p})-\Ip+\Iu}\right]\\ -{\Hp\,\Iu} \geq 0
\end{multline}
with equality if $\p=\p^{*}$. Hence, we need to minimize \eqref{eq.ratediff2} to obtain near optimal $\p$, i.e.,
\begin{equation}\label{eq.optts4} 
\begin{aligned}
&\underset{\p \in \mathbb{S}}{\text{minimize }}\quad
\ror(g)\left[{\Hp-{\Ip}+{\Iu}}\right]\\
&\phantom{\underset{\p \in \mathbb{S}}{\text{minimize }}\quad}-\Hp\,\Iu\\
& \text{subject to}
\quad \a^{T}\p-\beta_{\Delta}\,\Ip +\beta_{\Delta}\Hp -{P} \leq 0.
\end{aligned}
\end{equation}
Generally, the capacity $\ror(g)$  is not known a priori. Hence, we can substitute it with a preliminary estimate, $r\triangleq \rts(g,\Delta,\po)$, around an initial point $\po$ that is iteratively updated, as suggested in \cite{bocherer2012capacity}. After some manipulations and by dividing the objective function by the constant $\Iu$, the problem becomes
	\begin{subequations}\label{eq.optts5}
		\begin{alignat}{2}
		&\underset{\p \in \mathbb{S}}{\text{minimize }} &&\overbrace{r\left[{1-\frac{\Ip}{\Iu}}\right]}^{f_{0}(\p,r)}-\overbrace{\Hp\left[1-\frac{r}{\Iu} \right]}^{{\psi}_{0}(\p,r)}\\ 
		&\text{subject to} &&\ \overbrace{\a^{T}\p-\beta_{\Delta}\,\Ip-\!{P} }^{f_{1}(\p)}\,\,-\,\,\overbrace{\left(-\beta_{\Delta}\, \Hp\right)}^{{\psi}_{1}(\p)}\leq 0\, \label{eq.constopt5}.
		\end{alignat}
	\end{subequations}
The optimization problem \eqref{eq.optts5} is not convex, but it can be reformulated as a \ac{DC}  problem. In fact, both the objective function and   power constraint are represented as \ac{DC} functions, for $r>\Iu$ and  $\beta_{\Delta}\geq 0$ (i.e., $\Delta \geq 2\,P/[M-1]$). This is because $-\Ip$  and $-\Hp$  are convex functions, while $\a^{T}\p$ is an affine.

 A local minimum for the \ac{DC} problem can be obtained through many iterative algorithms, e.g., convex-concave procedure\cite{BoydLipp:16}. In this method, the subtracted convex function is approximated by a  Taylor expansion of the first-order around a feasible initial point, $\po$, which is successively updated till convergence.  More precisely, ${\psi}_{0}(\p,r)$ is replaced with
\begin{align}
{\bar \psi}_{0}(\p,\po,r) &\triangleq {\psi}_{0}(\po,r) +\sum\limits_{j=0}^{M-1} \frac{\delta\, {\psi}_{0}(\p,r)}{\delta\, \pj} {\bigg|}_{\pj=\poj} (\pj-\poj) \nonumber \\
&=\left[\frac{r}{\Iu}-1 \right] \sum\limits_{j=0}^{M-1} \pj\,\log(\poj)\,.
\end{align} 
Similarly, the constraint $f_{1}(\p)-{\psi}_{1}(\p)\leq 0$ can be convexified by substituting ${\psi}_{1}(\p)$ with ${\bar \psi}_{1}(\p,\po) \triangleq \beta_{\Delta}\textstyle \sum_{j=0}^{M-1} \pj\,\log(\poj)$. The convex-concave procedure is described in Algorithm~\ref{alg.1} for a fixed $\Delta$. Then, the golden section method can be used to search for the optimal constellation spacing, $\Delta^{*}$, and the associated probabilities $\p^{*} \triangleq {\p}_{{\Delta}^{*}}$  that lead to the maximum achievable rate.  
 Additionally, it can be proved that the solution of the convexified problem is still subject to the average power constraint in \eqref{eq.constopt5}. From the convexity of ${\psi}_{1}(\p)$, we have $ {\psi}_{1}(\p)\geq {\bar \psi}_{1}(\p,\po)$, leading to 
\begin{equation}
f_{1}(\p,r)-{\psi}_{1}(\p,\po,r) \leq f_{1}(\p,r)-\bar{\psi}_{1}(\p,\po,r) \leq 0\,.
\end{equation}
Although the solution is feasible,  some of the feasibility range is lost because of the power constraint relaxation. Hence, we can obtain a feasible sub-optimal input distribution that yields an achievable rate of the scheme with  \ac{SMD} metric. 
  \begin{algorithm}[t] 
  	\caption{\acs{coms} with Optimal \ac{FEC} Rate} 
  	\label{alg.1} 
  	\begin{algorithmic} [1]
  		\STATE\textbf{Input} 
  		$\Delta, {\po}$, $\delta_{1}$, $\delta_2\qquad$   \%{\small  $\delta_{1}$ and $\delta_{1}$ are the stopping criteria tolerances }
  		\REPEAT
  		\STATE $r:=\rts(g,\Delta,{\po})$
  		\REPEAT
  		\STATE Solve the following convex problem 
  		\begin{align*}
  		\poo= &\underset{\p\in \mathbb{S}}{\text{arg min}}
  		& & f_{0}(\p,r)-\bar{\psi}_{0}(\p,\po,r)\\
  		& \text{subject to}
  		& & f_{1}(\p)-\bar{\psi}_{1}(\p,\po) \leq 0
  		\end{align*}
  		\STATE $\delta_{f}:=[f_{0}(\po,r)-{\psi}_{0}(\po,r)]-[f_{0}(\poo,r)-{\psi}_{0}(\poo,r)] $
  		\STATE {\textbf{update}}	$\po:=\poo$
  		\UNTIL{$\left|\delta_{f}\right|<\delta_{1}$}
  		\UNTIL{$\left|\rts(g,\Delta,{\po})-r\right| <\delta_{2}$}
  		\STATE ${\p}_{\Delta}:=\po$
  		\STATE \textbf{Output} ${\p}_{\Delta}$, $\rts(g,\Delta,{\p}_{\Delta})$, $R_{\text{BMD}}(g,\Delta,{\p}_{\Delta})$
  	\end{algorithmic}
  \end{algorithm}

The capacity of the proposed scheme can be obtained through the above-indicated procedure; however, optimal \ac{FEC} encoders are required. More precisely,   the optimal channel coding rate, $\rc({\p}^{*})$, can take any real value between zero and one; however, most of the encoders allow only several operational modes with predefined \ac{FEC} rates. In this case, the system can sub-optimally operate on the maximum allowable \ac{FEC} rate, which is slightly less than $\rc({\p}^{*})$. This leads to a transmission rate that is lower than the achievable rate. Nevertheless, there is no guarantee that the average power constraint is not being violated  when operating on lower \ac{FEC} rate.  In the following section, we propose an approach to calculate the achievable rate of \ac{coms} with practical  \ac{BMD} with a finite set of coding rates.  
\section{The Achievable Rate of the Proposed \acs{coms} with Practical Channel Encoders}\label{sec.adaptivescheme}
In this section, we describe how the proposed adaptive scheme can be implemented using practical off-the-shelf channel encoders with predefined \ac{FEC} rates and \acp{BMD}. For example, the  \ac{DVB-S2} considers \ac{LDPC} channel encoders with coding rate
\begin{equation}\label{eq.fecrateset}
\rc\in {\mathbb R}_{\text{c}}=\left\{\frac{1}{4}, \frac{1}{3}, \frac{2}{5}, \frac{1}{2}, \frac{3}{5}, \frac{2}{3}, \frac{3}{4}, \frac{4}{5}, \frac{5}{6}, \frac{8}{9}, \frac{9}{10} \right\},
\end{equation}
while the newer standard DVB-S2X permits more rates\cite{DVB:14,Dvbs2x:15}. Then, the computational complexity of the proposed scheme is analyzed.  

\subsection{Capacity-Achieving Distribution}
We develop a method to obtain the input distribution that maximizes the  spectral efficiency of the proposed \ac{coms} with  \ac{BMD}. For a given \ac{FEC} rate, we would like to compute the maximum possible transmission rate that is achievable by the \ac{coms} scheme. The transmission rate is maximized, while satisfying both the power and rate constraints.  The maximum achievable transmission rate can be found by solving the following optimization problem 
\begin{equation}\label{eq.opttsfixedrc}
\begin{aligned}
& \underset{\Delta>0,\, \p\in \mathbb{S}}{\text{maximize}}
& &R(\p)\\
& \text{subject to}
& & \rc \,\a^{T}\p + (1- \rc) \Delta\, \frac{M-1}{2}\,  \leq {P},\\
&&&R(\p) \leq \rts(g,\Delta,{\mathbf p})-\rb,
\end{aligned}
\end{equation}
where $\rb \leq (1-\rc)\Iu$ is a back-off rate to account for the reduced achievable rate with \ac{BMD}, which can be iteratively set as indicated in Algorithm~\ref{alg.2}.\footnote{ The back-off rate can further be set to compensate for the rate loss due to the use of  finite-length channel encoders. }
The three main differences  between the optimization problem in \eqref{eq.opttsfixedrc} when compared to \eqref{eq.optts2} is that the transmission rate is maximized rather than the mutual information,  the rate is forced to be less than the mutual information with a margin allowing the use of a low complexity  \ac{BMD}, and the \ac{FEC} rate is fixed. These modifications allow obtaining the maximum achievable rate of the  \ac{coms} scheme with practical \ac{FEC} encoders and \ac{BMD}. Unfortunately, the optimization problem \eqref{eq.opttsfixedrc} is not convex, because of the second constraint. Nevertheless,  it can be convexified by noting that the constraint can be reformulated as a \ac{DC} problem. Then, the convex-concave procedure is applied as before. More precisely, we iteratively solve the following convex problem  
\begin{equation}\label{eq.opttsfixedrc2}
\begin{aligned}
&\underset{\Delta>0,\, \p\in \mathbb{S}}{\text{maximize}} & & \rc \, \Hp\\
& \text{subject to}
& & \rc \,\a^{T}\p + (1- \rc)\Delta\,\frac{M-1}{2}\,  \leq {P},\\
&&&-\rc\,\Ip-(1-\rc)\,\Iu \\
&&&\phantom{-} - \rc\, \textstyle \sum\limits_{j=0}^{M-1} \pj\,\log(\poj)+\rb\leq 0,
\end{aligned}
\end{equation}
where $\po$ is an initial feasible point, as shown in Algorithm~\ref{alg.2}. The optimal constellation spacing is obtained also using the golden section search method. The procedure is repeated for each supported \ac{FEC} rate, and the optimal channel coding rate $c^{*} \in {\mathbb R}_{\text{c}}$ is the one that yields the maximum transmission rate for the proposed scheme, denoted by $R_{\text{\acs{coms}}}(g)$. 

The same procedure can be repeated for all the modulation orders $M$ supported by the encoder. Then, the $M$-ary modulation that yields the largest transmission rate, for the considered \ac{SNR}, is selected. Alternatively, in order to reduce the modulation complexity, we can opt for the minimum $M$-ary modulation that yields the target rate for the considered \ac{SNR}. 

The proposed scheme adapts the rate according to the channel condition. Hence, the ergodic transmission rate for the \ac{coms}  scheme can be written as
\begin{equation}\label{eq.ergodicrate}
{\bar{R}_{\text{\acs{coms}}}}\left(\pw/\sigma\right)=\int_{0}^{\infty} R_{\text{\acs{coms}}}(g)\, f_{G}(g)\,\d g,
\end{equation} 
where $R_{\text{\acs{coms}}}(g)$ is the maximum transmission rate for a given instantaneous \ac{SNR}, i.e., $g \pw/\sigma$,  and $f_{G}(g)$ is the \ac{PDF} of the irradiance in \eqref{eq.pdfgammagamma}.  
\begin{algorithm}[t] 
	\caption{\acs{coms} with Practical \ac{FEC} Encoders} 
	\label{alg.2} 
	\begin{algorithmic}[1] 
		\STATE\textbf{Input} 
		$\Delta, {\po}$ , $\delta,\rb \qquad$   \% {\small $\delta$ is the stopping criteria tolerance }
		\REPEAT
		\REPEAT
		\STATE {$\rb:=\min\left(\rb,(1-\rc)\Iu\right) $}
		\STATE Solve the convex problem \eqref{eq.opttsfixedrc2} with  interior-point method
		\STATE Assign the optimal \ac{PMF} to $\poo$
		\STATE $\delta_{R}:=R(\poo)-R(\po)$
		\STATE\textbf{Update}	$\po:=\poo$
		\UNTIL{$\left|\delta_{R}\right|<\delta$}
		\STATE $\p_{\Delta}^{*}:=\po$
		\STATE  $\delta_{\text{BMDRp}} := R_{\text{BMD}}(g,\Delta,\p_{\Delta}^{*})-R(\p_{\Delta}^{*}) $  \% {\small The  transmission rate back-off with respect to the \ac{BMD} }
		
		\IF{$\delta_{\text{BMDRp}} < 0$}  
	
		\STATE {$\rb:=\rb+\delta_{\text{BMDRp}} $}
		\ENDIF
		\UNTIL{$\delta_{\text{BMDRp}} \geq 0$}
		\STATE \textbf{Output} $\p_{\Delta}^{*}$,  $\rts(g,\Delta,\p_{\Delta}^{*})$
		, $R_{\text{BMD}}(g,\Delta,\p_{\Delta}^{*})$
	\end{algorithmic}
\end{algorithm}

%
\subsection{Computational Complexity of the Proposed Scheme}
In the following, the computational complexity of the proposed scheme is analyzed.  The complexity is mainly due to the  numerical optimization, required for computing the input distribution, and  distribution matching/dematching through the \ac{CCDM}.\footnote{The \ac{FEC}  coding stage is standard for both uniform and non-uniform signaling; hence, the associated complexity is not discussed here.}
For computing the capacity-achieving distribution of the proposed scheme, Algorithm~\ref{alg.2} should be run for each value of $\Delta$. Let us define  $z_{\delta}$  as the number of iterations till rate convergence in Algorithm~\ref{alg.2}, which depends on the stopping criteria $\delta$,  $z_{\text{GS}}$ as the number of iterations till convergence for the golden section algorithm to find the optimal $\Delta$, and $z_{\rc}$ as the number of code rates supported by the \ac{FEC} coder. Hence, in order to find the optimal distribution, we need to solve  $z_{\delta}\,z_{\text{GS}}\, z_{\rc}$ convex optimization problems. The  computation complexity for solving a single convex optimization problem using interior-point method is on the order of $M^3$, i.e., $\mathcal{O}(M^{3})$ \cite{PotWri:20,GolMic:89}. 

The optimization can either be performed online or offline, depending on the available computational capability  at the transmitter. For instance, considering offline optimization,  the optimal modulation order, probabilities of symbols, constellation spacing, and \ac{FEC}  rate can be obtained for a predefined set of \acp{SNR} with fine arbitrary quantization and stored in the memory, reducing the computational complexity at the expense of some rate loss due to the quantization.

For the \ac{DM}, the \ac{CCDM}, implemented using  arithmetic coding \cite{SchBoc:16}, requires $k_{\text p}$  iterations for the matching, while it needs $\n_{\text p}$ for dematching. Each iteration involves  $M$ additions, multiplications, and comparisons \cite{GulFehalv:19}. Unfortunately,  the algorithms for arithmetic coding are sequential in nature; hence, it is a challenging task to parallelize the implementation \cite{FehMilKoiPar:19}. 

In the literature, several methods have been proposed to  reduce the computational complexity and increase the parallelization capability of \ac{DM} methods, including \ac{CCDM} \cite{BocSteSch:17,PikWenKramer:19,FehMilKoiPar:19,FehMilKoiPars:20}. In particular, efficient implementation of the \ac{CCDM} using finite-precision algorithms has been proposed in \cite{PikWenKramer:19}. Also,   a parallel architecture for a \ac{CCDM} has been proposed with a subset ranking algorithm rather than arithmetic codes \cite{FehMilKoiPars:20}.     Alternatively, the \ac{MPDM} is a non-constant composition \ac{DM} that can achieve lower  rate loss compared to \ac{CCDM} for a given $n_{\text p}$ \cite{FehMilKoiPar:19}. However, the output sequences have different empirical distributions that match the target distribution only on average, i.e., the ensemble average of the output sequences imitates the desired distribution.  Hence, the empirical distribution of a specific sequence may deviate from the desired distribution. 

Besides \ac{DM} algorithms, there are other methods for \ac{PS} with indirect signal shaping algorithms (e.g., sphere shaping and shell mapping) that opt for the most energy-efficient codewords \cite{WilWui:93,LarFarTre:94}. On the contrary to \ac{DM} methods that target a specific pre-designed distribution in the output, the indirect methods are designed to achieve a target rate, i.e., a constant ${k_{p}}$ for a given $n_{\text{p}}$. For example, in shell mapping, $2^{k_{p}}$ codewords are chosen from all possible sequences that fulfill the energy budget, and the others are neglected. The potential codewords are on the surface or inside an $n_{\p}$-dimensional sphere\cite{ForGal:84}. The computational complexity of enumerative sphere shaping \cite{WilWui:93}  and shell mapping  \cite[Algorithm 2]{LarFarTre:94}  is $\mathcal{O}(n_{\text{p}})$ and $\mathcal{O}({n_{\text{p}}}^3)$, respectively \cite{GulFehalv:19}.  The rate loss for indirect signal shaping can be smaller than \ac{CCDM}; however, the rate loss as a function of the sequence length is not the only parameter to judge the performance of the \ac{PS} algorithms.  In fact, the \ac{CCDM} can afford a longer frame length with lower computational complexity compared to shell mapping \cite{ChoWinzer:19}. Therefore, it could be preferable to use \ac{CCDM} over sphere mapping for longer frame lengths, even if the \ac{CCDM} rate loss is higher for a fixed frame length. Another issue for sphere shaping and shell mapping is that they are not optimized for \ac{IM/DD} optical communications. This is attributed to the fact that shell and sphere mapping are asymptotically optimal for \ac{AWGN} channels with electrical power constraint (restricted signal variance)\cite{ForGal:84}. On the other hand,  the input distribution for \ac{IM/DD} is subject to non-negativity and optical power constraints (restricted signal mean); hence, a sphere is not necessarily the optimal shape for the shell in this case.

It may be of interest also to investigate the power consumption of the proposed scheme. In fact, the power dissipation can be attributed to two primary sources: the communication and computation power consumption. The proposed scheme reduces the communication power consumption, as it can achieve the same rate compared to uniform signaling with less transmitted power (up to $2$ dB). The reason is that the \ac{CCDM} is asymptotically optimal from a power efficiency perspective \cite{ChoWinzer:19}. Nevertheless, the power dissipated in computations increases due to the additional operations required for the \ac{CCDM}. The savings in the communication power  increases significantly with the distance, while the computational power dissipation is distance-independent. Hence, the proposed scheme is energy efficient for backhauling applications. However, a careful analysis is required to precisely judge the total power consumption,  accounting for the considered hardware components, e.g., the microprocessor and power amplifier,  their impact on the actual dissipated energy for each arithmetic operation, and the savings in the transmitted power \cite{Tucker:11,ElzGioChisyndrome:19}.
\section{Outage Probability due to the Irradiance Fluctuations}\label{sec.outage}
The outage probability of the proposed \ac{coms} is the probability that a given transmission rate is not achievable, because of the irradiance fluctuations. The outage probability can be written as
\begin{equation}\label{eq.outageprob}
P_{\text{outage}}(R)=\mathbb{P}\left\{ R_{\text{BMD}}(g,\Delta,\p)<R(\p)\right\},
\end{equation}
where $R$ is the transmission rate. In the proposed \ac{coms}  when the \ac{CSI}  is available at the transmitter, the encoder adapts the transmission rate according to the channel condition $g$. From the rate constraint  \eqref{eq.opttsfixedrc}, the rate is always achievable with \ac{SMD} and optimal \ac{FEC}, even with zero back-off rate. For \ac{BMD} and practical \ac{FEC}, an appropriate $\rb$ in \eqref{eq.opttsfixedrc} should be considered to achieve an acceptable error performance. 

Regarding the proposed blind \ac{coms} when the \ac{CSI} is known only at the receiver, there is a non-zero outage probability. In order to calculate it we start by defining the threshold for the irradiance
\begin{equation}\label{eq.goutage}
{g}_{\text{o}}(R)=\left\{{g}_{\text{o}} \in \mathbb{R}^{+} :  R_{\text{BMD}}\left(g_{\text{o}},\Delta,\p\right)= R(\p)              \right\},
\end{equation}
where $\mathbb{R}^{+}$ is the set of all positive real numbers.
The achievable rate $ R_{\text{BMD}}\left(g_{\text{o}},\Delta,\p\right)$ is non-decreasing in $g$. Hence,  from  \eqref{eq.outageprob}, \eqref{eq.goutage}, and \cite[Eq.~11]{NisTsi:09}, the outage probability can be rewritten in closed-form  as
\begin{multline}\label{eq.outageprobnocsi}
P_{\text{outage}}(R)=\mathbb{P}\left\{ G<{g_{\text{o}}(R)}\right\}=\frac{\left(\alpha\, \beta\, {g}_{\text o}(R)\right)^{\frac{\alpha+\beta}{2}} }{\Gamma (\alpha) \Gamma (\beta)}\\
\times G_{1,3}^{2,1}\left(\displaystyle{\alpha\, \beta\, {g}_{\text o}(R)}\left|
\begin{array}{c}
\displaystyle{ 1-\frac{\alpha+\beta}{2}} \\
\displaystyle{\frac{\alpha-\beta}{2},\frac{\beta-\alpha}{2},-\frac{\alpha+\beta}{2} }
\end{array}
\right.\right),
\end{multline}
where $G_{1,3}^{2,1}\left(\cdot|\cdot\right)$ is the Meijer G-function\cite{GraRyz:B01}.

For the blind \ac{coms}, we propose to design the encoder such that the outage probability is less than a predefined threshold $\bar{\gamma}$. In this regard, the transmission rate is calculated for a fixed irradiance $\bar{g}(\bar{\gamma})$ defined as
\begin{eqnarray}
\bar{g}(\bar{\gamma}) =\left\{\bar{g} \in \mathbb{R}^{+} :   \mathbb{P}\left\{ G< \bar{g}\right\}=    \bar{\gamma}                \right\},
\end{eqnarray}
which can be found from the inverse of the turbulence fading \acl{CDF}. Finally, the transmission rate is obtained for a given $\pw/\sigma$ by substituting $g$ with $\bar{g}$ in the procedures indicated in  section~\ref{sec.adaptivescheme}.  Consequently, the outage probability is bounded below $\bar{\gamma}$ as required, i.e.,  $P_{\text{outage}}(R_{\text{\acs{coms}}}(\bar{g}))\leq \bar{\gamma}$.
\section{Numerical results}\label{sec.numericalresults}
\begin{figure}
	\centering
	\includegraphics[width=0.992\linewidth,clip]{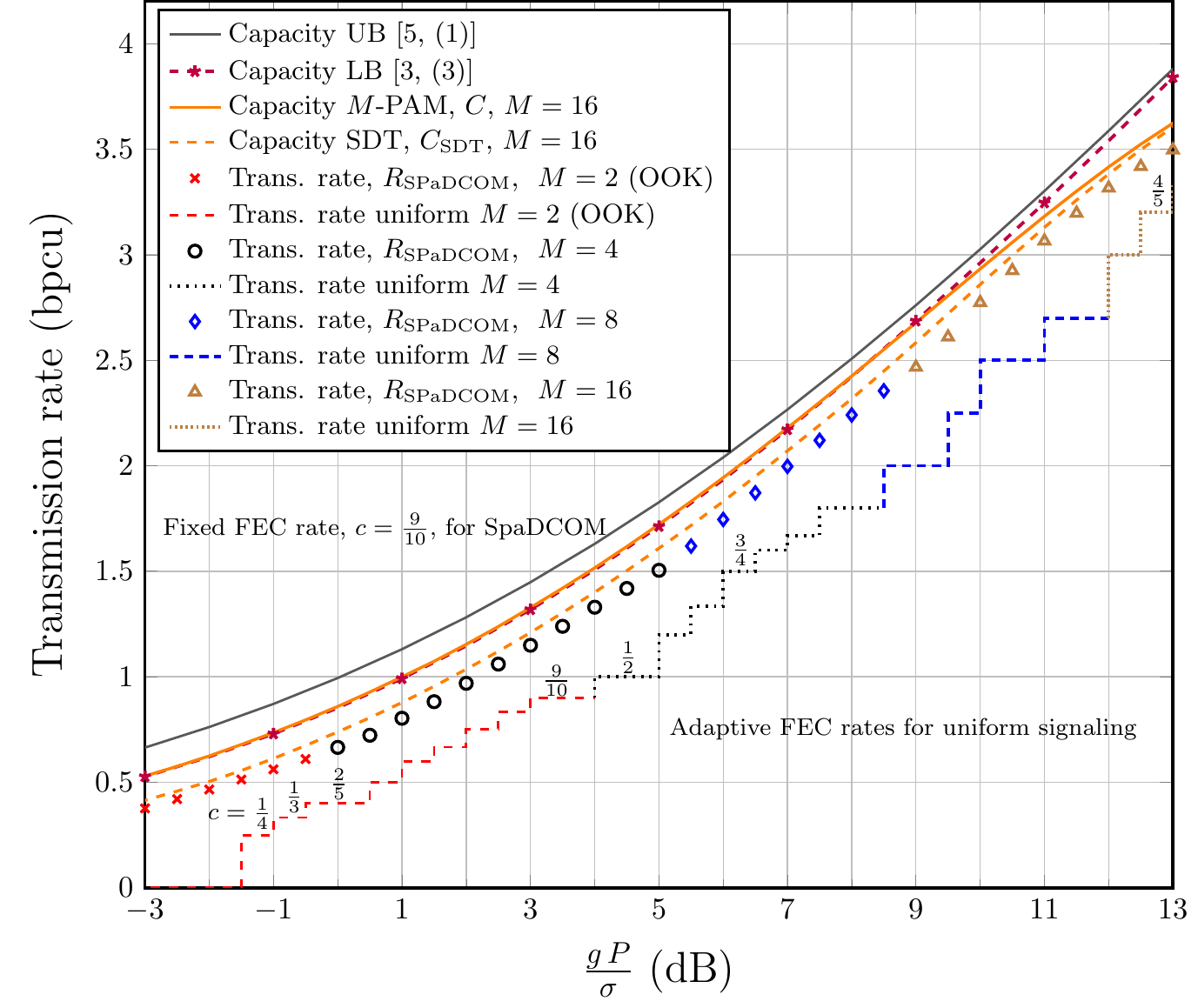}
	\caption{The  transmission rate vs \ac{SNR} for the \ac{coms} adaptive scheme in Section~\ref{sec.adaptivescheme} with $\rb=0.05$, for various modulation orders $M\in\{2,4,8,16 \}$, along with the transmission rate of uniform signaling,  achievable rate of $16$-\acs{PAM}, and capacity bounds in 	\cite{ChaMorSlim:16} and \cite{FaridHran:10}.}
	\label{fig:FixedRatemaxEntropy}
\end{figure}
In this section,  Monte Carlo simulations and numerical results are depicted to evaluate the performance of the proposed \ac{coms} schemes. The uniform signaling, \ac{MPAM} capacity,  capacity bounds in 	\cite[(1)]{ChaMorSlim:16} and \cite[(3)]{FaridHran:10}, and coded modulation scheme in \cite{HeBoKim:19} are used as benchmarks for the performance. In all numerical results, the optimal distribution for the proposed scheme is obtained according to the procedures  in \cref{sec.adaptivescheme}. The  back-off rate, required input parameter for Algorithm~\ref{alg.2},  is set as $\rb=0.05$.  For the channel coding, we consider the \ac{LDPC} \ac{DVB-S2} with $\rc\in {\mathbb R}_{\text{c}}$, defined in \cref{sec.adaptivescheme}.

In \cref{fig:FixedRatemaxEntropy}, the transmission rate of the proposed \ac{coms} with \ac{CSI} versus \ac{SNR} is compared with the achievable rate of the uniform signaling and the capacity of  $16$-\acsu{PAM}. {\color{black} The  modulation order $M$ is adapted with the \ac{SNR} for both uniform and non-uniform signaling schemes}. The $16$-\ac{PAM} is chosen as a benchmark, as it can be considered as an upper bound for the capacities of lower $M$-ary modulations. The optimal rate that maximizes the achievable rate for the \ac{coms} tends to use the highest \ac{FEC} rate, i.e., $\rc=0.9$, to approach the \ac{MPAM} capacity. On the other hand, the \ac{FEC} rate for the uniform based scheme is selected such that the transmission rate, $\rc\,\log(M)$, is less than the achievable rate with uniform signaling, $\Iu$. {\color{black} For instance, at $R=1.5$, the proposed scheme operates within $1.75$ from the capacity upper bound, $1$~dB from both the \ac{MPAM} capacity and the capacity lower bound, and within $0.3$ dB from the \ac{SDT} capacity.} Also, it outperforms the uniform signaling with more than $1$~dB for $R=1.5$, and up to $2.5$~dB for $R=0.5$, where the gap increases for lower transmission rates. 
\begin{figure}[t]
	\centering
	\begin{subfigure}{0.5\textwidth}
		\centering
		\includegraphics[width=0.992\linewidth,clip]{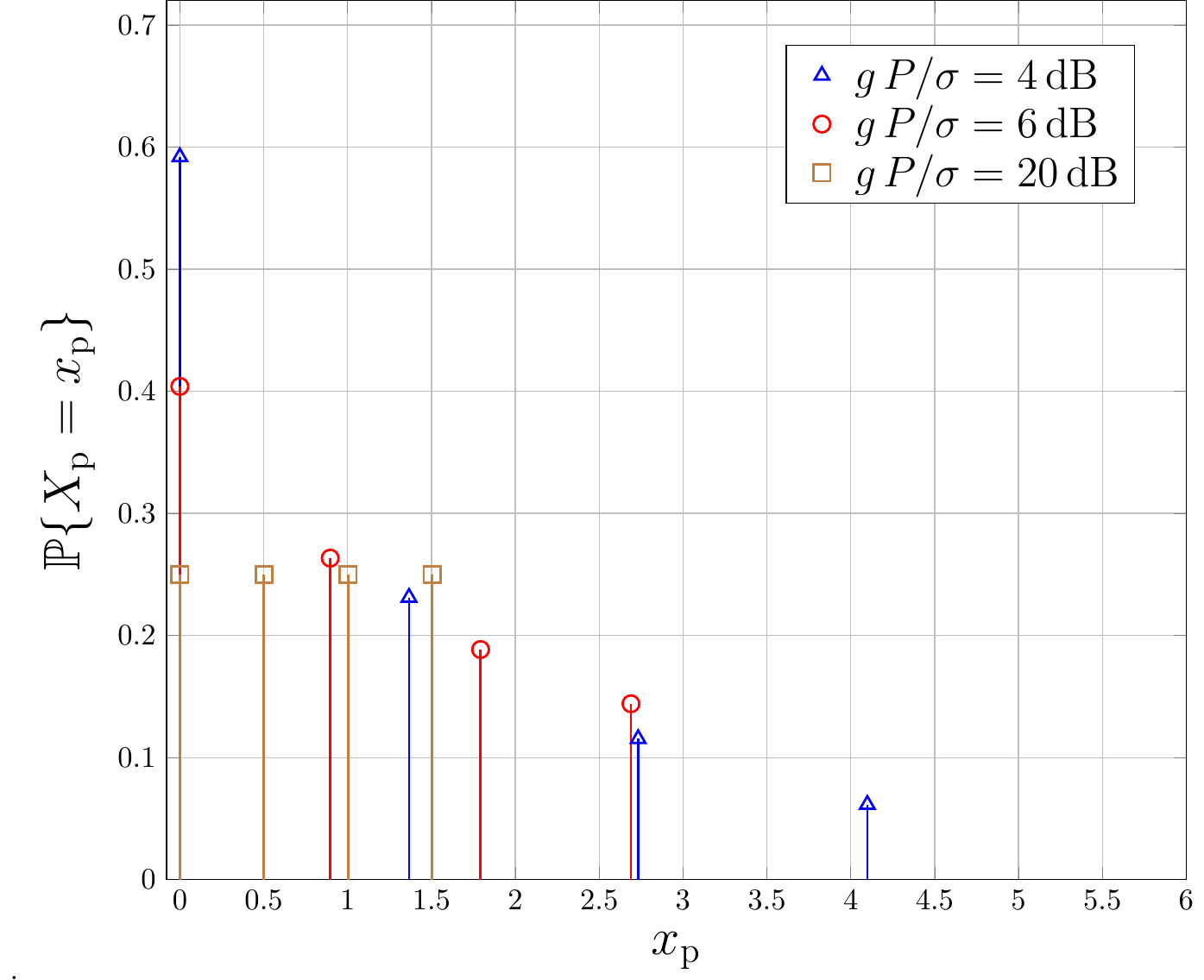}
		\caption{}
		\label{fig:pmf1}
	\end{subfigure}%
	~
	
	\begin{subfigure}{0.5\textwidth}
		\centering
		\includegraphics[width=0.992\linewidth,clip]{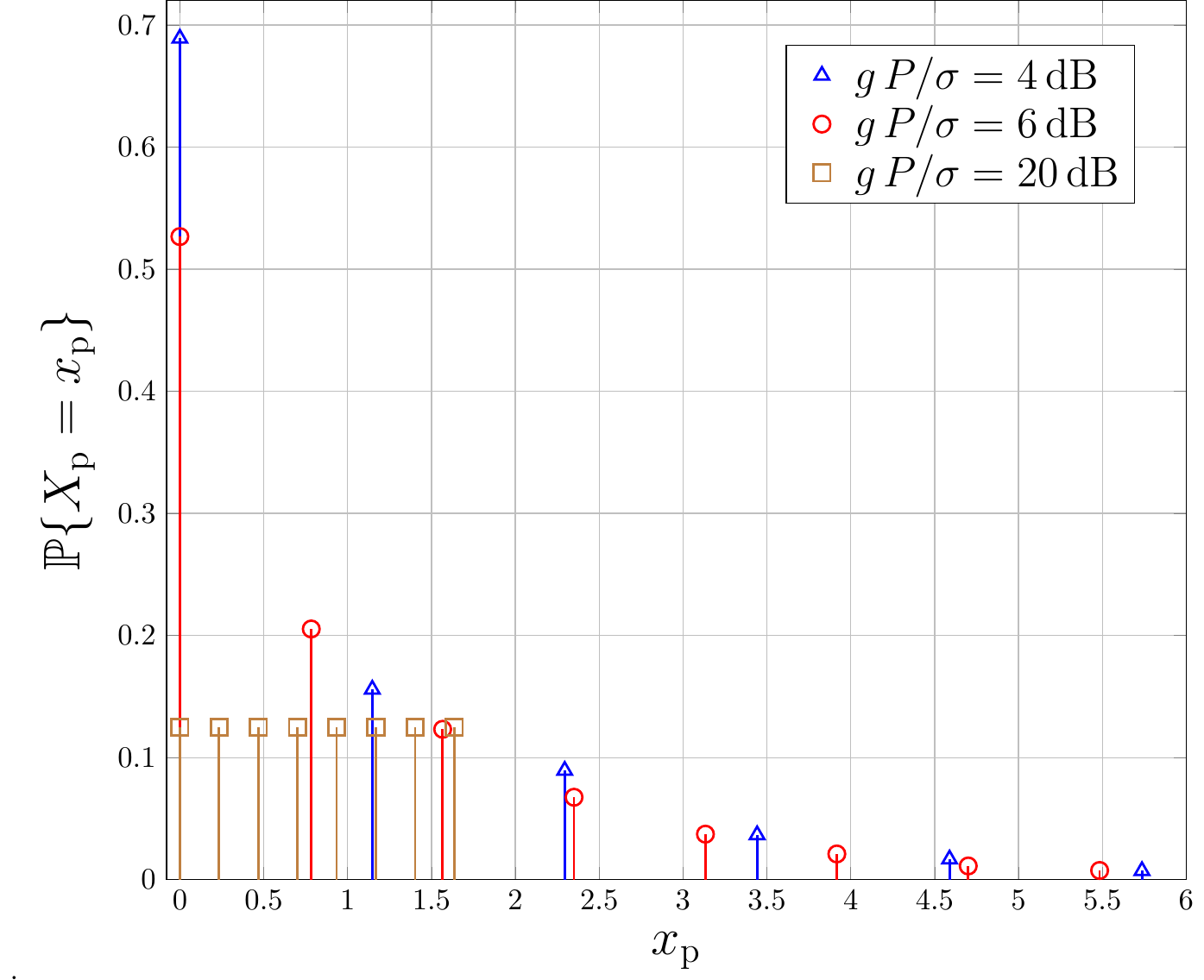}
		\caption{}
		\label{fig:pmf2}
	\end{subfigure}
	\caption{The optimal \acl{PMF} of the input signal for various instantaneous \acp{SNR}, obtained following the procedures in  \cref{sec.adaptivescheme}: (a) $M=4$, and (b) $M=8$.}
	\label{fig:PMF}
\end{figure}%
In \cref{fig:PMF}, the optimal \ac{PMF} obtained following the procedures in \cref{sec.adaptivescheme} is shown for various \acp{SNR} and modulation orders  $M\in \{4,8\}$.  It can be noticed that the symbols with low amplitudes tend to have higher probabilities than the symbols with larger amplitudes at low \acp{SNR}. This permits large constellation spacing $\Delta^{*}$ without violating the average optical power constraint. For instance, the maximum value of $\Delta$ in \ac{OOK} increases with the probability of the zero symbol, i.e.,  $\Delta \leq {\pw}/{(1-p_{0})}$.
For high instantaneous \acp{SNR} and a  fixed $M$, the distribution is almost uniform, and the distance between the symbols is small. This is attributed to the fact that for asymptotically high  \ac{SNR} the mutual information approaches the source entropy that is maximized using equiprobable symbols.\footnote{Note that the proposed adaptive scheme should increase the modulation order $M$ to achieve higher rates at high \acp{SNR}.}

\begin{figure}[t!]
	\centering
	\includegraphics[width=0.992\linewidth]{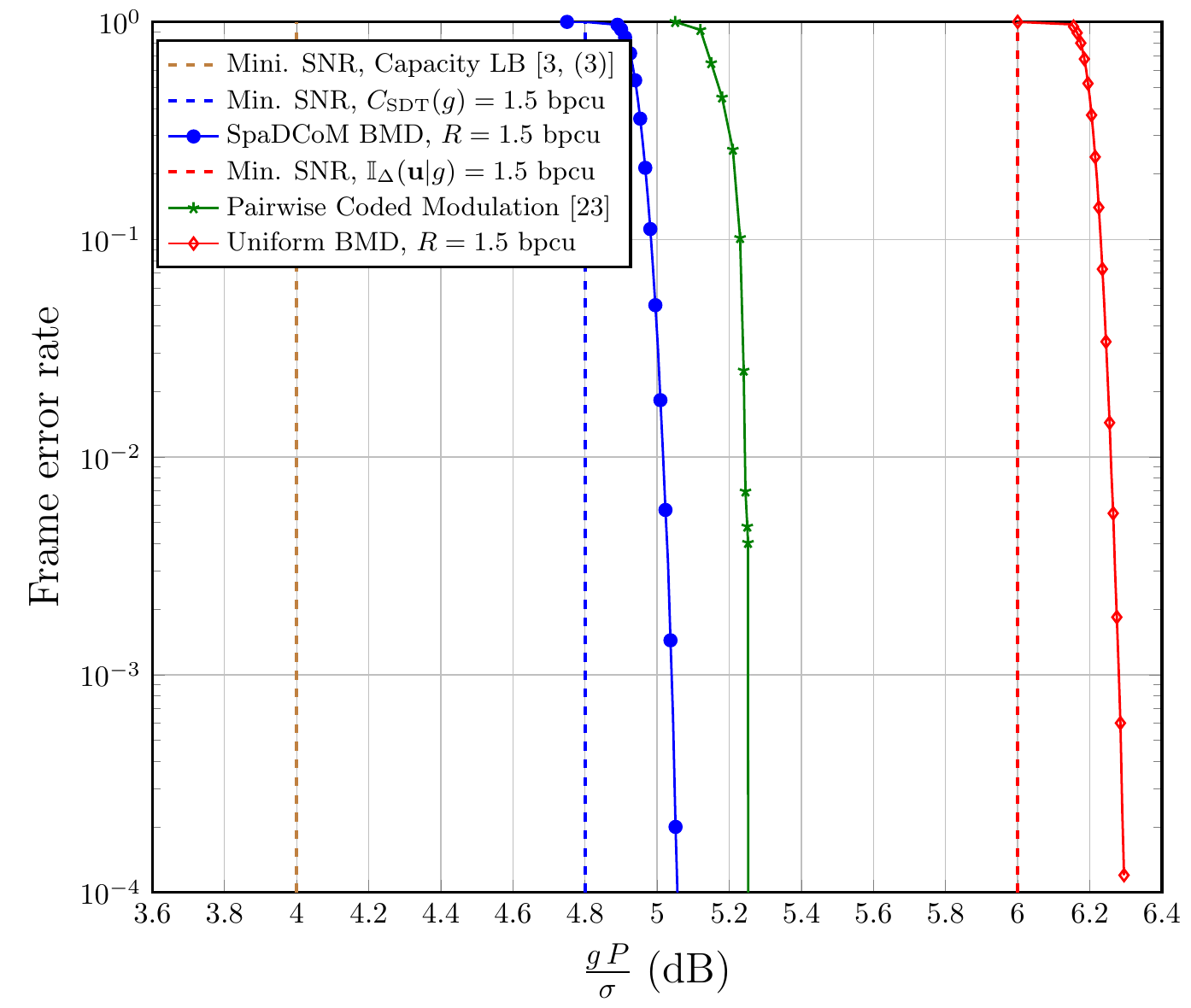}
	\caption{The frame error rate vs \ac{SNR} of the proposed \ac{coms} scheme,  uniform signaling, and pairwise   \ac{CM} scheme \cite{HeBoKim:19}, for $M=4$ and transmission rate $R=1.5$~bpcu. The channel coding rate of  uniform signaling is $\rc=0.75$, while $\rc=0.9$ for both   \ac{coms} and pairwise  \ac{CM}.}
	\label{fig:ber}
\end{figure}
In \cref{fig:ber}, the performance of the proposed scheme is compared with  both uniform signaling  and pairwise \acl{CM} \cite{HeBoKim:19},  in terms of the \ac{FER} using Monte Carlo simulation. The transmission rates of the schemes are kept constant at $1.5$~bpcu. The minimum \ac{SNR} that permits transmission at this rate can be found from Fig.~\ref{fig:m4achievablea} for the capacity lower bound  \cite[(3)]{FaridHran:10}, \ac{SDT},  and uniform  based scheme as $4$~dB, $4.8$~dB,  and $6$~dB, respectively. For \ac{coms}, the minimum \ac{SNR} that is required for an achievable rate of $1.5$~bpcu is $5$~dB, as shown in Fig.~\ref{fig:FixedRatemaxEntropy}. The optimal symbol probabilities, constellation spacing, and \ac{FEC} rate are  $\p^{*}=[0.53,0.25,0.14,0.08]$, $\Delta^{*}=1.18$, and $\rc=9/10$, respectively. These parameters  are obtained using the procedures indicated in \cref{sec.adaptivescheme} with $\rb=0.05$ in   Algorithm~\ref{alg.2}. The corresponding values for  uniform signaling are $\p=[1/4,1/4,1/4,1/4]$, $\Delta_{\text{u}} \triangleq 2 \pw/ (M-1)$, and  $\rc=3/4$, respectively. As an additional benchmark, the performance of the \ac{CM} scheme in \cite{HeBoKim:19} is shown. The \ac{FEC} rate and the pairwise distribution, i.e., two consecutive symbols have the same probability, are adjusted to yield the target rate and power. In particular, we set the pairwise \ac{PMF}  as $[0.405,0.405,0.095,0.095]$ with $1.14$ constellation spacing, while $\rc=0.9$.   The \ac{LDPC} code adopted for \ac{DVB-S2}  is considered for all the schemes  with word length $64800$.  For the decoding of the proposed scheme, the  \ac{LLR} is computed  from \eqref{eq.LLR} assuming perfect estimation of the channel gain $g$. 
We can see that the \ac{FER} exhibits a phase transition phenomenon around the \ac{SNR}, which is associated with the transmission rate. Also, the proposed scheme achieves about $1.3$~dB and $0.25$~dB reduction in transmitted power compared to  uniform and pairwise signaling, respectively, for \ac{FER}$=10^{-3}$ and $R=1.5$.  

\begin{figure}[t!]
	\centering
	\begin{subfigure}{0.5\textwidth}
		\centering
		\includegraphics[width=0.992\linewidth]{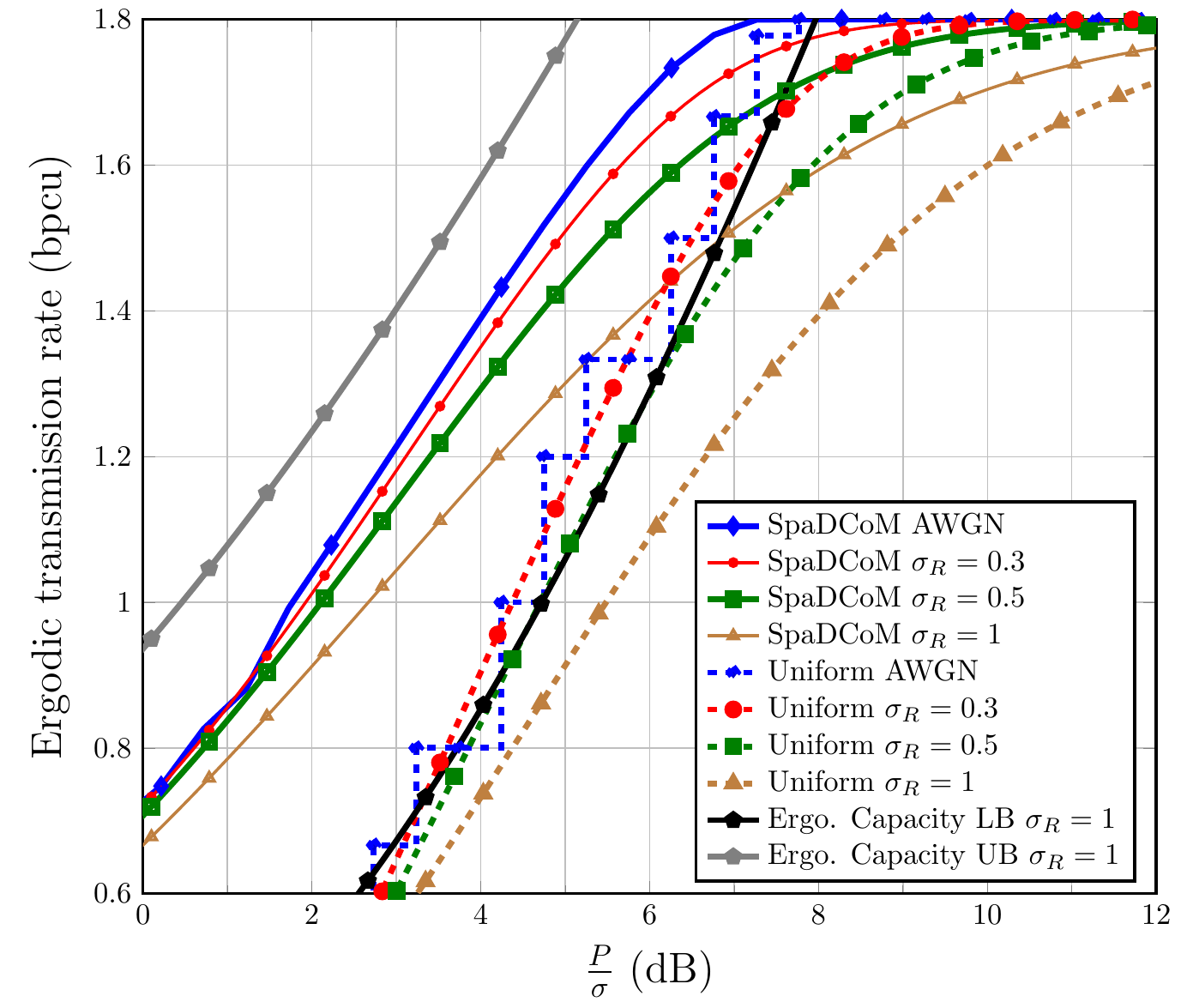}
		\caption{} 
	\end{subfigure}%
	~
	
	\begin{subfigure}{0.5\textwidth}
		\centering
		\includegraphics[width=0.992\linewidth]{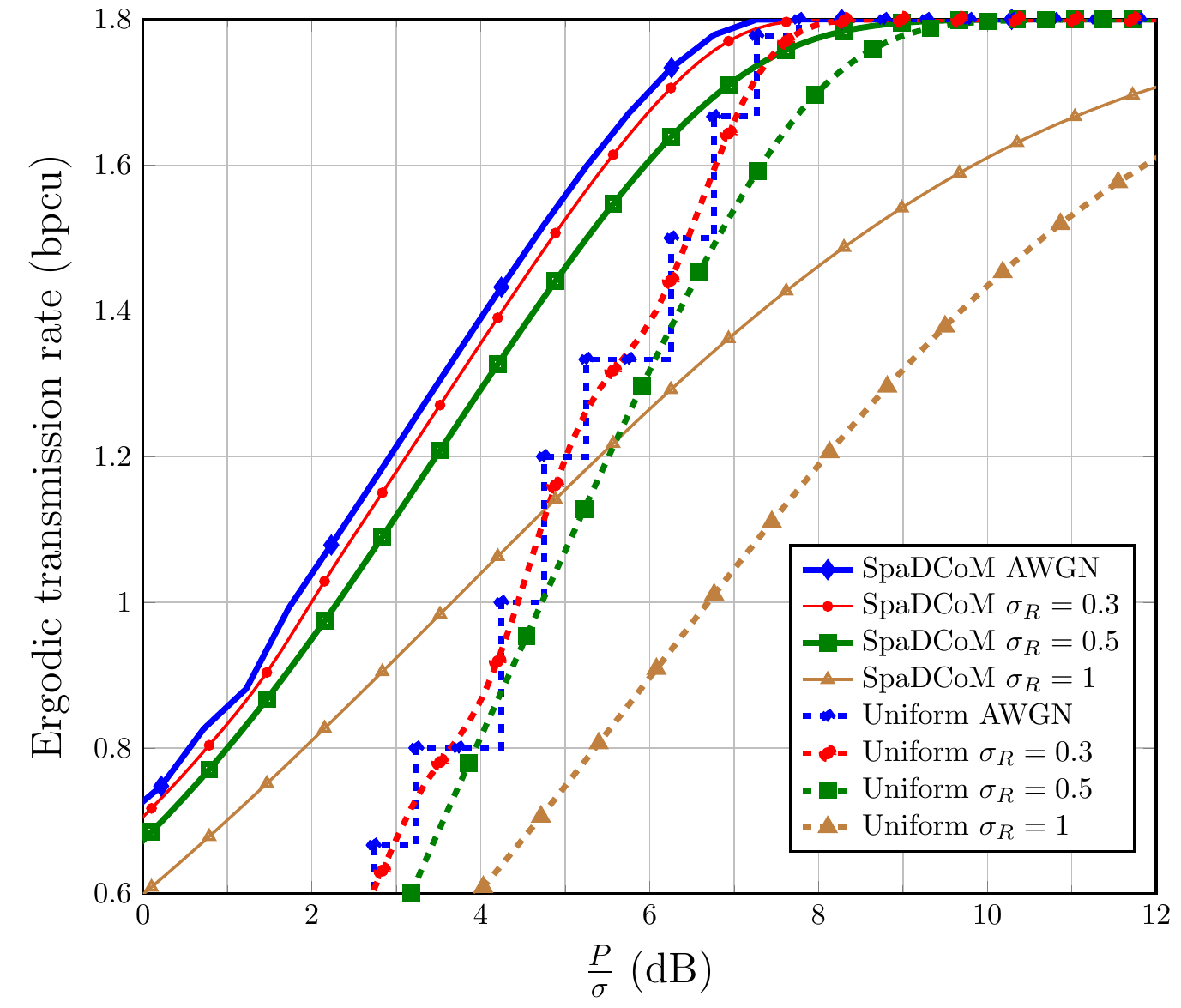}
		\caption{}
	\end{subfigure}
	\caption{The ergodic rate of the \ac{CSI}-aware scheme vs $\pw/\sigma$, for $M=4$, and various turbulence conditions in terms of the Rytov variance, when the irradiance fluctuations are modeled by:
		 (a) Gamma-Gamma   (b) Lognormal distributions. }
	\label{fig:ergodicrate}
\end{figure}
\begin{figure}
	\centering
	\includegraphics[width=0.992\linewidth]{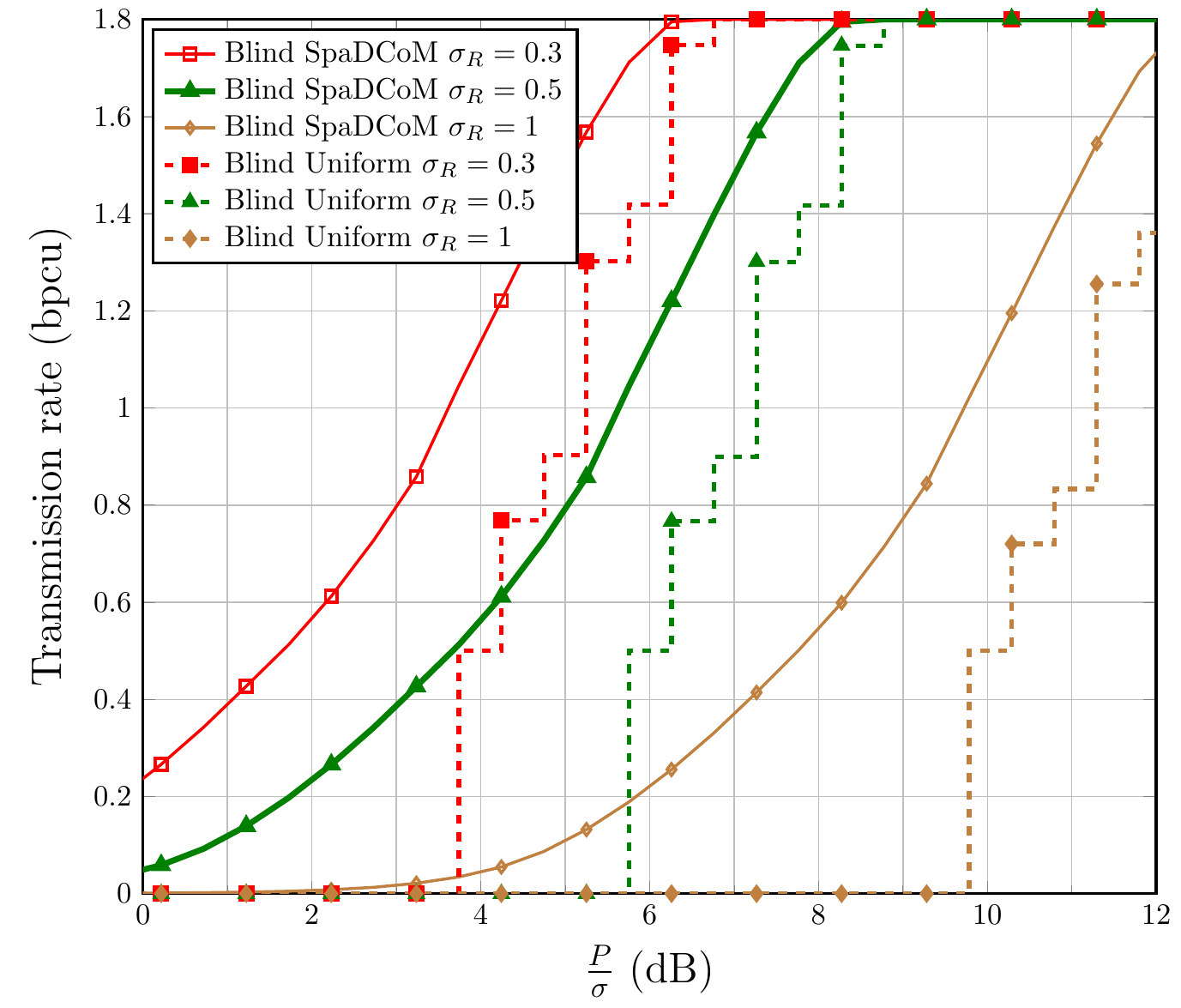}
	\caption{The transmission rate of the blind \ac{coms} scheme versus \ac{SNR} with outage probability $P_{\text{outage}} \leq 10^{-4}$ and $M=4$.}
	\label{fig:fixedoutager}
\end{figure}
The ergodic rates of the \ac{coms} \ac{CSI} aware scheme, \eqref{eq.ergodicrate}, and the uniform approach are depicted in \cref{fig:ergodicrate} with $M=4$, for various turbulence conditions in terms of the Rytov variance.\footnote{Note that the ergodic rate for uniform signaling in turbulence  is  continuous instead of the stairwise rate as in  \ac{AWGN} channels because the ergodic rate is  the average with respect to the continuous \ac{r.v.} $G$ representing the channel gain.} As a benchmark, we calculate ergodic upper and lower bounds on the capacity from the bounds in \cite{LapMoser:09}. For comparison, we also consider the lognormal model for the atmospheric turbulence-induced fading in \cite[Eq.~(34)]{Andrews:01}. The proposed scheme achieves better ergodic performance compared to the uniform signaling based method, e.g., $2.5$~dB at $R=0.56$. It can be seen that as the turbulence increases (i.e., the Rytov variance $\sigma^{2}_{\text{R}}$), the ergodic rate decreases. Also, for small turbulence with $\sigma_{\text{R}}=0.1$, the ergodic rate approaches the transmission rate for \ac{AWGN} channels, while the gap between the rate and the ergodic upper bound of the capacity in \cite{LapMoser:09} is about $2$~dB at $R=1.2$~bpcu.

Finally, the performance of the blind \ac{coms} scheme when \ac{CSI} is known only at the decoder is illustrated in Fig.~\ref{fig:fixedoutager}. The transmission rate of the proposed scheme versus $\pw/\sigma$ is compared with the uniform signaling for  different turbulence conditions with Gamma-Gamma distributed turbulence. The design criteria is according to \cref{sec.outage} with upper bounded outage probability such that $P_{\text{outage}} \leq \bar{\gamma}= 10^{-4}$. For the proposed scheme, we set the transmission rate as $\rc=0.9$.  The \ac{FEC} rate for  uniform signaling is chosen such that the transmission rate, $\rc\,\log(M)$, is less than the achievable rate with uniform signaling, $R_{\text{BMD}}\left(\bar{g}(\bar{\gamma}),\Delta_{\text{u}} ,\u\right)$. The \ac{coms} encoder is superior to the uniform method with about $2$~dB for $R=0.5$~\ac{bpcu}, and with $1$~dB for $R=1.5$ and $\sigma_{\text R}=0.5$. 
\section{Conclusion}\label{sec.conclusion}
In this paper, we propose a coded modulation scheme for \acl{FSO} based backhaul applications. The encoder is adaptive to the atmospheric turbulence-induced fading with arbitrary fine granularity. In particular, the signal constellation is probabilistically shaped by a low complexity fixed-to-fixed length distribution matcher to approach the capacity of  \ac{FSO} channels with \ac{IM/DD}. The proposed method can employ any efficient off-the-shelf \ac{FEC} encoder, and it can also operate when the \ac{CSI} is known only at the receiver. The proposed scheme approaches the capacity of unipolar \ac{MPAM}. Moreover, it outperforms the uniform signaling based encoders. For instance, the probabilistic based scheme can achieve a reduction in the transmitted power up to $2$~dB compared to the uniform signaling at an ergodic rate of $0.5$~bpcu.   
\bibliographystyle{IEEEtran}
\bibliography{IEEEabrv,SignalShapingBib}

\begin{thebibliography}{10}
\providecommand{\url}[1]{#1}
\csname url@samestyle\endcsname
\providecommand{\newblock}{\relax}
\providecommand{\bibinfo}[2]{#2}
\providecommand{\BIBentrySTDinterwordspacing}{\spaceskip=0pt\relax}
\providecommand{\BIBentryALTinterwordstretchfactor}{4}
\providecommand{\BIBentryALTinterwordspacing}{\spaceskip=\fontdimen2\font plus
\BIBentryALTinterwordstretchfactor\fontdimen3\font minus
  \fontdimen4\font\relax}
\providecommand{\BIBforeignlanguage}[2]{{%
\expandafter\ifx\csname l@#1\endcsname\relax
\typeout{** WARNING: IEEEtran.bst: No hyphenation pattern has been}%
\typeout{** loaded for the language `#1'. Using the pattern for}%
\typeout{** the default language instead.}%
\else
\language=\csname l@#1\endcsname
\fi
#2}}
\providecommand{\BIBdecl}{\relax}
\BIBdecl

\bibitem{AlzShakYanAlouini:18}
M.~{Alzenad}, M.~Z. {Shakir}, H.~{Yanikomeroglu}, and {M. -S. Alouini},
  ``{FSO}-based vertical backhaul/fronthaul framework for {5G}+ wireless
  networks,'' \emph{{IEEE} Commun. Mag.}, vol.~56, no.~1, pp. 218--224, Jan
  2018.

\bibitem{KhaUysal:14}
M.~A. {Khalighi} and M.~{Uysal}, ``Survey on free space optical communication:
  A communication theory perspective,'' \emph{{IEEE} Commun. Surveys Tuts.},
  vol.~16, no.~4, pp. 2231--2258, Fourthquarter 2014.

\bibitem{FaridHran:10}
A.{Farid} and S.~{Hranilovic}, ``Capacity bounds for wireless optical intensity
  channels with {Gaussian} noise,'' \emph{{IEEE} Trans. Inf. Theory}, vol.~56,
  no.~12, pp. 6066--6077, Dec 2010.

\bibitem{LapMoser:09}
A.~Lapidoth, S.~M. Moser, and M.~A. Wigger, ``On the capacity of free-space
  optical intensity channels,'' \emph{{IEEE} Trans. Inf. Theory}, vol.~55,
  no.~10, pp. 4449--4461, 2009.

\bibitem{ChaMorSlim:16}
A.~{Chaaban}, J.~{Morvan}, and {M. -S. Alouini}, ``Free-space optical
  communications: Capacity bounds, approximations, and a new sphere-packing
  perspective,'' \emph{{IEEE} Trans. Commun.}, vol.~64, no.~3, pp. 1176--1191,
  March 2016.

\bibitem{ChaRezSlim:17}
A.~{Chaaban}, Z.~{Rezki}, and {M. -S. Alouini}, ``Fundamental limits of
  parallel optical wireless channels: Capacity results and outage
  formulation,'' \emph{{IEEE} Trans. Commun.}, vol.~65, no.~1, pp. 296--311,
  Jan 2017.

\bibitem{ChaRezkiSlim:17}
------, ``Low-{SNR} capacity of parallel {IM-DD} optical wireless channels,''
  \emph{{IEEE} Commun. Lett.}, vol.~21, no.~3, pp. 484--487, March 2017.

\bibitem{ChaRezSlim:16}
------, ``On the capacity of the {Intensity-Modulation Direct-Detection}
  optical broadcast channel,'' \emph{{IEEE} Trans. Wireless Commun.}, vol.~15,
  no.~5, pp. 3114--3130, May 2016.

\bibitem{ChaAlEbraNafSlim:17}
A.~{Chaaban}, O.~M.~S. {Al-Ebraheemy}, T.~Y. {Al-Naffouri}, and {M. -S.
  Alouini}, ``Capacity bounds for the {Gaussian IM-DD} optical multiple-access
  channel,'' \emph{{IEEE} Trans. Wireless Commun.}, vol.~16, no.~5, pp.
  3328--3340, May 2017.

\bibitem{ForGal:84}
G.~{Forney}, R.~{Gallager}, G.~{Lang}, F.~{Longstaff}, and S.~{Qureshi},
  ``Efficient modulation for band-limited channels,'' \emph{{IEEE} J. Sel.
  Areas Commun.}, vol.~2, no.~5, pp. 632--647, Sep. 1984.

\bibitem{Ungerboeck:87}
G.~Ungerboeck, ``{Trellis}-coded modulation with redundant signal sets part
  {I}: Introduction,'' \emph{{IEEE} Commun. Mag.}, vol.~25, no.~2, pp. 5--11,
  1987.

\bibitem{GoldsmithChua:98}
A.~J. Goldsmith and S.-G. Chua, ``Adaptive coded modulation for fading
  channels,'' \emph{{IEEE} Trans. Commun.}, vol.~46, no.~5, pp. 595--602, 1998.

\bibitem{Djordjevic:May10}
I.~B. Djordjevic, ``Adaptive modulation and coding for free-space optical
  channels,'' \emph{J. Opt. Commun. Netw.}, vol.~2, no.~5, pp. 221--229, May
  2010.

\bibitem{ChaLioKar:11}
N.~D. {Chatzidiamantis}, A.~S. {Lioumpas}, G.~K. {Karagiannidis}, and
  S.~{Arnon}, ``Adaptive subcarrier {PSK} intensity modulation in free space
  optical systems,'' \emph{{IEEE} Trans. Commun.}, vol.~59, no.~5, pp.
  1368--1377, May 2011.

\bibitem{San:11}
H.~G. {Sandalidis}, ``Coded free-space optical links over strong turbulence and
  misalignment fading channels,'' \emph{{IEEE} Trans. Commun.}, vol.~59, no.~3,
  pp. 669--674, March 2011.

\bibitem{AnguitaDjordjevic:05}
J.~Anguita, I.~Djordjevic, M.~Neifeld, and B.~Vasic, ``Shannon capacities and
  error-correction codes for optical atmospheric turbulentchannels,'' \emph{J.
  Opt. Netw.}, vol.~4, no.~9, pp. 586--601, Sep 2005.

\bibitem{BocSteSch:15}
G.~{Böcherer}, F.~{Steiner}, and P.~{Schulte}, ``Bandwidth efficient and
  rate-matched low-density parity-check coded modulation,'' \emph{{IEEE} Trans.
  Commun.}, vol.~63, no.~12, pp. 4651--4665, Dec 2015.

\bibitem{BucSte:15}
F.~{Buchali}, G.~{Böcherer}, W.~{Idler}, L.~{Schmalen}, P.~{Schulte}, and
  F.~{Steiner}, ``Experimental demonstration of capacity increase and
  rate-adaptation by probabilistically shaped 64-{QAM},'' in \emph{Euro. Conf.
  on Opt. Comm. (ECOC)}, Sep. 2015, pp. 1--3.

\bibitem{BocSch:19}
G.~{Böcherer}, P.~{Schulte}, and F.~{Steiner}, ``Probabilistic shaping and
  forward error correction for fiber-optic communication systems,'' \emph{J. of
  Lightwave Technology}, vol.~37, no.~2, pp. 230--244, Jan 2019.

\bibitem{GitMatSte:19}
A.~D. Git, B.~Matuz, and F.~Steiner, ``Protograph-based {LDPC} code design for
  probabilistic shaping with on-off keying,'' in \emph{Annual Conf. on Inform.
  Sci. and Sys. (CISS), Baltimore, Maryland, USA}, March. 2019, pp. 1--6.

\bibitem{FarHar:09}
A.~{Farid} and S.~{Hranilovic}, ``Channel capacity and non-uniform signalling
  for free-space optical intensity channels,'' \emph{{IEEE} J. Sel. Areas
  Commun.}, vol.~27, no.~9, pp. 1553--1563, December 2009.

\bibitem{ChoWinzer:19}
J.~{Cho} and P.~J. {Winzer}, ``Probabilistic constellation shaping for optical
  fiber communications,'' \emph{J. of Lightwave Technology}, vol.~37, no.~6,
  pp. 1590--1607, March 2019.

\bibitem{HeBoKim:19}
Z.~He, T.~Bo, and H.~Kim, ``Probabilistically shaped coded modulation for
  {IM/DD} system,'' \emph{Opt. Express}, vol.~27, no.~9, pp. 12\,126--12\,136,
  Apr. 2019.

\bibitem{Bocherer:14}
G.~Bocherer, ``Probabilistic signal shaping for bit-interleaved coded
  modulation,'' in \emph{Proc. IEEE Inter. Symp. on Inf. Theory}.\hskip 1em
  plus 0.5em minus 0.4em\relax Citeseer, 2014.

\bibitem{YanZibLar:14}
M.~P. {Yankov}, D.~{Zibar}, K.~J. {Larsen}, L.~P.~B. {Christensen}, and
  S.~{Forchhammer}, ``Constellation shaping for fiber-optic channels with {QAM}
  and high spectral efficiency,'' \emph{{IEEE} Photon. Technol. Lett.},
  vol.~26, no.~23, pp. 2407--2410, Dec 2014.

\bibitem{PanKsc:16}
C.~{Pan} and F.~R. {Kschischang}, ``Probabilistic 16-{QAM} shaping in {WDM}
  systems,'' \emph{J. of Lightwave Technology}, vol.~34, no.~18, pp.
  4285--4292, Sep. 2016.

\bibitem{SheLiva:18}
A.~{Sheikh}, A.~G. i.~{Amat}, G.~{Liva}, and F.~{Steiner}, ``Probabilistic
  amplitude shaping with hard decision decoding and staircase codes,'' \emph{J.
  of Lightwave Technology}, vol.~36, no.~9, pp. 1689--1697, May 2018.

\bibitem{HuYanDa:18}
H.~{Hu}, M.~P. {Yankov}, F.~{Da Ros}, Y.~{Amma}, Y.~{Sasaki}, T.~{Mizuno},
  Y.~{Miyamoto}, M.~{Galili}, S.~{Forchhammer}, L.~K. {Oxenløwe}, and
  T.~{Morioka}, ``Ultrahigh-spectral-efficiency {WDM/SDM} transmission using
  {PDM-1024-QAM} probabilistic shaping with adaptive rate,'' \emph{J. of
  Lightwave Technology}, vol.~36, no.~6, pp. 1304--1308, March 2018.

\bibitem{AlhaAndPhi:01}
R.~L.~P. Ammar Al-Habash, Larry C.~Andrews, ``Mathematical model for the
  irradiance probability density function of a laser beam propagating through
  turbulent media,'' \emph{Optical Engineering}, vol.~40, no.~8, pp. 1554 --
  1562 -- 9, 2001.

\bibitem{BenRezkiSlim:13}
F.~Benkhelifa, Z.~Rezki, and M.-S. Alouini, ``{Low SNR capacity of {FSO} links
  over Gamma-Gamma atmospheric turbulence channels},'' \emph{{IEEE} Commun.
  Lett.}, vol.~17, no.~6, pp. 1264--1267, 2013.

\bibitem{HraKac:04}
S.~{Hranilovic} and F.~R. {Kschischang}, ``Capacity bounds for power- and
  band-limited optical intensity channels corrupted by gaussian noise,''
  \emph{{IEEE} Trans. Inf. Theory}, vol.~50, no.~5, pp. 784--795, 2004.

\bibitem{CoverThomas:06}
T.~Cover and J.~Thomas, \emph{Elements of information theory}, 2nd~ed.\hskip
  1em plus 0.5em minus 0.4em\relax John Wiley \& Sons, 2006.

\bibitem{BoydVan:04}
S.~Boyd and L.~Vandenberghe, \emph{Convex optimization}.\hskip 1em plus 0.5em
  minus 0.4em\relax Cambridge university press, 2004.

\bibitem{SchBoc:16}
P.~{Schulte} and G.~{Böcherer}, ``Constant composition distribution
  matching,'' \emph{{IEEE} Trans. Inf. Theory}, vol.~62, no.~1, pp. 430--434,
  Jan 2016.

\bibitem{BocSteSch:17}
G.~{Böcherer}, F.~{Steiner}, and P.~{Schulte}, ``Fast probabilistic shaping
  implementation for long-haul fiber-optic communication systems,'' in
  \emph{Proc. {Euro.} Conf. on Opt. Comm. (ECOC)}, 2017, pp. 1--3.

\bibitem{PikWenKramer:19}
M.~Pikus, W.~Xu, and G.~Kramer, ``Finite-precision implementation of arithmetic
  coding based distribution matchers,'' \emph{arXiv preprint arXiv:1907.12066},
  2019.

\bibitem{FehMilKoiPar:19}
T.~{Fehenberger}, D.~S. {Millar}, T.~{Koike-Akino}, K.~{Kojima}, and
  K.~{Parsons}, ``Multiset-partition distribution matching,'' \emph{{IEEE}
  Trans. Commun.}, vol.~67, no.~3, pp. 1885--1893, Nov 2019.

\bibitem{FehMilKoiPars:20}
------, ``Parallel-amplitude architecture and subset ranking for fast
  distribution matching,'' \emph{{IEEE} Trans. Commun.}, vol. Preprint, pp.
  1--1, Jan 2020.

\bibitem{AbrSte:72}
M.~Abramowitz and I.~A. Stegun, \emph{Handbook of mathematical functions with
  formulas, graphs, and mathematical tables}, 9th~ed.

\bibitem{DVB:14}
``Digital video broadcasting ({DVB}): Second generation framing structure,
  channel coding and modulation systems for broadcasting, interactive services,
  news gathering and other broadband satellite applications,'' Euro. Telecomm.
  Stand. Institute, EN, Tech. Rep. 302 307-1, V1. 4.1, 2014.

\bibitem{Mackay:99}
D.~J.~C. {MacKay}, ``Good error-correcting codes based on very sparse
  matrices,'' \emph{{IEEE} Trans. Inf. Theory}, vol.~45, no.~2, pp. 399--431,
  March 1999.

\bibitem{MarFab:09}
A.~{Martinez}, A.~{Guillen i Fabregas}, G.~{Caire}, and F.~M.~J. {Willems},
  ``Bit-interleaved coded modulation revisited: {A} mismatched decoding
  perspective,'' \emph{{IEEE} Trans. Inf. Theory}, vol.~55, no.~6, pp.
  2756--2765, June 2009.

\bibitem{Boch:14}
G.~{Böcherer}, ``Probabilistic signal shaping for bit-metric decoding,'' in
  \emph{IEEE Inter. Symp on Inf. Theory}, June 2014, pp. 431--435.

\bibitem{AlvAgrLavMah:15}
A.~{Alvarado}, E.~{Agrell}, D.~{Lavery}, R.~{Maher}, and P.~{Bayvel},
  ``Replacing the soft-decision {FEC} limit paradigm in the design of optical
  communication systems,'' \emph{J. of Lightwave Technology}, vol.~33, no.~20,
  pp. 4338--4352, Oct 2015.

\bibitem{ChoSchWin:17}
J.~{Cho}, L.~{Schmalen}, and P.~J. {Winzer}, ``Normalized generalized mutual
  information as a forward error correction threshold for probabilistically
  shaped {QAM},'' in \emph{Proc. Euro. Conf. on Opt. Comm. (ECOC)}, Sep. 2017,
  pp. 1--3.

\bibitem{AlvFehCheWil:18}
A.~Alvarado, T.~Fehenberger, B.~Chen, and F.~M.~J. Willems, ``Achievable
  information rates for fiber optics: Applications and computations,'' \emph{J.
  Lightwave Technol.}, vol.~36, no.~2, pp. 424--439, Jan 2018.

\bibitem{YosAl:19}
T.~Yoshida, A.~Alvarado, M.~Karlsson, and E.~Agrell, ``{Post-FEC BER}
  prediction for bit-interleaved coded modulation with probabilistic shaping,''
  \emph{arXiv preprint arXiv:1911.01585}, 2020.

\bibitem{bocherer2012capacity}
G.~B{\"o}cherer, ``{Capacity-Achieving Probabilistic Shaping for Noisy and
  Noiseless Channels},'' {Ph.D. Dissertation}, Hochschulbibliothek der
  Rheinisch-Westf{\"a}lischen Technischen Hochschule Aachen, 2012.

\bibitem{BoydLipp:16}
T.~Lipp and S.~Boyd, ``Variations and extension of the convex--concave
  procedure,'' \emph{Optimization and Engineering}, vol.~17, no.~2, pp.
  263--287, Jun. 2016.

\bibitem{Dvbs2x:15}
``{White paper on the use of DVB-S2X for DTH applications, DSNG \& professional
  services, broadband interactive services and VL-SNR applications},'' Digital
  Video Broadcasting, Tech. Rep., March 2015.

\bibitem{PotWri:20}
F.~A. Potra and S.~J. Wright, ``Interior-point methods,'' \emph{Journal of
  Computational and Applied Mathematics}, vol. 124, no.~1, pp. 281 -- 302,
  2000, numerical Analysis 2000. Vol. IV: Optimization and Nonlinear Equations.

\bibitem{GolMic:89}
D.~Goldfarb and M.~J. Todd, ``Chapter {II} linear programming,''
  \emph{Handbooks in Operations Research and Management Science}, vol.~1, pp.
  73--170, 1989.

\bibitem{GulFehalv:19}
Y.~C. G{\"u}ltekin, T.~Fehenberger, A.~Alvarado, and F.~M. Willems,
  ``Probabilistic shaping for finite blocklengths: {Distribution} matching and
  sphere shaping,'' \emph{arXiv preprint arXiv:1909.08886}, 2019.

\bibitem{WilWui:93}
F.~Willems and J.~Wuijts, ``A pragmatic approach to shaped coded modulation,''
  in \emph{Proc. IEEE 1st Symp. on Commun. and Veh. Technol. in the
  Benelux}.\hskip 1em plus 0.5em minus 0.4em\relax Citeseer, 1993.

\bibitem{LarFarTre:94}
R.~{Laroia}, N.~{Farvardin}, and S.~A. {Tretter}, ``On optimal shaping of
  multidimensional constellations,'' \emph{{IEEE} Trans. Inf. Theory}, vol.~40,
  no.~4, pp. 1044--1056, Jul 1994.

\bibitem{Tucker:11}
R.~S. {Tucker}, ``Green optical communications—{Part I}: {Energy} limitations
  in transport,'' \emph{IEEE Journal of Selected Topics in Quantum
  Electronics}, vol.~17, no.~2, pp. 245--260, 2011.

\bibitem{ElzGioChisyndrome:19}
A.~{Elzanaty}, A.~{Giorgetti}, and M.~{Chiani}, ``Lossy compression of noisy
  sparse sources based on syndrome encoding,'' \emph{{IEEE} Trans. Commun.},
  vol.~67, no.~10, pp. 7073--7087, Oct. 2019.

\bibitem{NisTsi:09}
H.~Nistazakis, T.~Tsiftsis, and G.~Tombras, ``Performance analysis of
  free-space optical communication systems over atmospheric turbulence
  channels,'' \emph{IET comm.}, vol.~3, no.~8, pp. 1402--1409, 2009.

\bibitem{GraRyz:B01}
I.~Gradshteyn and I.~Ryzhik, \emph{Tables of Integrals, Series, and Products},
  6th~ed.\hskip 1em plus 0.5em minus 0.4em\relax San Diego, CA: Academic Press,
  Inc., 1994.

\bibitem{Andrews:01}
L.~C. Andrews, R.~L. Phillips, and C.~Y. Hopen, \emph{{Laser Beam Scintillation
  with Applications}}.\hskip 1em plus 0.5em minus 0.4em\relax Bellingham,
  Washington 98227-0010 USA: SPIE—The International Society for Optical
  Engineering, 2001.

\end{thebibliography}
\vspace{3cm}
\begin{IEEEbiography}[{\includegraphics[width=1in,height=1.2in,clip]{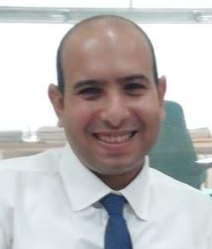}}]{Ahmed Elzanaty}(S’13-M’19) received the Ph.D. degree (excellent cum laude) in Electronics, Telecommunications, and Information technology from the University of Bologna, Italy, in 2018.  He was a research fellow at the University of Bologna from 2017 to 2019. Currently, he is a post-doctoral fellow at King Abdullah University of Science and Technology (KAUST), Saudi Arabia. He has participated in several national and European projects, such as GRETA and EuroCPS. His research interests include  coded modulation, compressive sensing, and distributed training of neural networks. He is the recipient of the best paper award at the IEEE Int. Conf. on Ubiquitous Wireless Broadband (ICUWB 2017). 
\end{IEEEbiography}
\vspace{-15.0 cm}
\begin{IEEEbiography}[{\includegraphics[width=1.11in,clip]{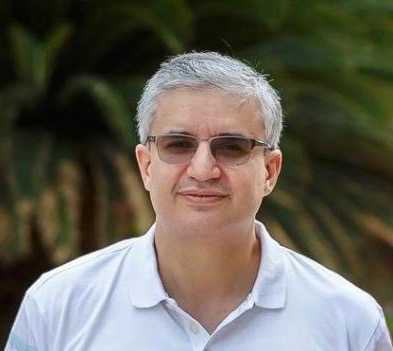}}]{Mohamed-Slim Alouini}
	(S'94-M'98-SM'03-F'09) was born in Tunis, Tunisia. He received the Ph.D. degree in Electrical Engineering
	from the California Institute of Technology (Caltech), Pasadena, CA, USA, in 1998. He served as a faculty member in the University of Minnesota,
	Minneapolis, MN, USA, then in the Texas A\&M University at Qatar, Education City, Doha, Qatar before joining King Abdullah University of
	Science and Technology (KAUST), Thuwal, Makkah Province, Saudi Arabia as a Professor of Electrical Engineering in 2009. His current research interests include the modeling, design, and performance analysis of wireless communication systems.
\end{IEEEbiography}

\end{document}